\documentclass[twocolumn]{aastex631}

\usepackage{amsmath}	

\newcommand*\diff{\mathop{}\!\mathrm{d}}

\newcommand{\newsuggestion}[1]

\usepackage[utf8x]{inputenc}
\usepackage{url}
\shorttitle{tSZ-ISW cross-correlation}
\shortauthors{Ibitoye et al.}

\graphicspath{{./}{figures/}}

\begin{document}

\title{Cross-correlation between the Thermal Sunyaev\textendash Zeldovich Effect and the Integrated Sachs\textendash Wolfe Effect}

\author[0000-0002-0966-8598]{Ayodeji Ibitoye}
\affiliation{National Astronomical Observatories, Chinese Academy of Sciences, 
\\20A Datun Road, Chaoyang District, Beijing 100101, P. R. China}
\affiliation{Department of Physics and Electronics, Adekunle Ajasin University, P. M. B. 001,\\ Akungba-Akoko, Ondo State, South-West Nigeria}

\author{Wei-Ming Dai}
\affiliation{School of Physical Science and Technology, Ningbo University, Ningbo 315211, China}

\author[0000-0001-8108-0986]{Yin-Zhe Ma}
\thanks{Corresponding author: Y.-Z. Ma, \url{mayinzhe@sun.ac.za}}
\affiliation{Department of Physics, Stellenbosch University, Matieland, Western Cape, 7602, South Africa}
\affiliation{National Institute for Theoretical and Computational
Sciences (NITheCS), South Africa}

\author[0000-0003-0051-272X]{Patricio Vielva}
\affiliation{Instituto de Física de Cantabria (CSIC - UC), Avda. Los Castros s/n, E-39005 Santander, Spain}

\author{Denis Tramonte}
\affiliation{Department of Physics, Xi'an Jiaotong-Liverpool University, 111 Ren'ai Road, \\    \quad Suzhou Dushu Lake Science and Education Innovation District, Suzhou Industrial Park, Suzhou 215123, P.R. China}

\author[0000-0001-5475-2919]{Amare Abebe}
\affiliation{Centre for Space Research, North-West University, Potchefstroom 2520, South Africa}
\affiliation{National Institute for Theoretical and Computational
Sciences (NITheCS), South Africa}

\author{Aroonkumar Beesham}
\affiliation{Department of Mathematical Sciences, University of Zululand, Private Bag X1001, Kwa-Dlangezwa 3886, South Africa}
\affiliation{National Institute for Theoretical and Computational
Sciences (NITheCS), South Africa}
\affiliation{Faculty of Natural Sciences, Mangosuthu University
of Technology, P.O. Box 12363, Jacobs 4052, South Africa}

\author{Xuelei Chen}
\affiliation{National Astronomical Observatories, Chinese Academy of Sciences, 
\\20A Datun Road, Chaoyang District, Beijing 100101, P. R. China}

\begin{abstract}
We present a joint cosmological analysis of the power spectra measurement of the Planck Compton parameter and the integrated Sachs\textendash Wolfe (ISW) maps. We detect the statistical correlation between the Planck Thermal Sunyaev\textendash Zeldovich (tSZ) map and ISW data with a significance of a $3.6\sigma$ confidence level~(CL), with the autocorrelation of the Planck tSZ data being measured at a
$25 \sigma$ CL. The joint auto- and cross-power spectra constrain the matter density to be $\Omega_{\rm m}= 0.317^{+0.040}_{-0.031}$, the Hubble constant $H_{0}=66.5^{+2.0}_{-1.9}\,{\rm km}\,{\rm s}^{-1}\,{\rm Mpc}^{-1}$ and the rms matter density fluctuations to be $\sigma_{8}=0.730^{+0.040}_{-0.037}$ at the 68\% CL. The  derived large-scale structure $S_{8}$ parameter is $S_8 \equiv \sigma_{8}(\Omega_{\rm m}/0.3)^{0.5} = 0.755\pm{0.060} $. If using only the diagonal blocks of covariance matrices, the Hubble constant becomes $H_{0}=69.7^{+2.0}_{-1.5}\,{\rm km}\,{\rm s}^{-1}\,{\rm Mpc}^{-1}$. In addition, we obtain the constraint of the product of the gas bias, gas temperature, and density as $b_{\rm gas} \left(T_{\rm e}/(0.1\,{\rm keV})  \right ) \left(\bar{n}_{\rm e}/1\,{\rm m}^{-3} \right) = 3.09^{+0.320}_{-0.380}$. We find that this constraint leads to an estimate on the electron temperature today as $T_{\rm e}=(2.40^{+0.250}_{-0.300}) \times 10^{6} \,{\rm K}$, consistent with the expected temperature of the warm\textendash hot intergalactic medium. Our studies show that the ISW\textendash tSZ cross-correlation is capable of probing the properties of the large-scale diffuse gas.
\end{abstract}

\keywords{galaxies: clusters: general - galaxies: clusters: power spectrum - cosmic infrared background - cosmological parameters - cosmology: large scale structure of Universe}


\section{Introduction} 
\label{sec:introduction} 

Understanding the composition and distribution of baryonic matter in the Universe is a crucial step toward unraveling the mysteries of its formation and evolution. While baryons make up a small fraction of the total matter\textendash energy content~\citep{ACTDR4_20,Planck2018TT}, they play a vital role in the processes that shape the cosmic structures that we observe today. Early estimates and numerical simulations show that most baryons are ``missing,'' whereas the baryons that are already made into stars and galaxies constitute a small portion of the total baryon budget~\citep{Fukugita1998,Fukugita04,cenostriker2006,Shull}. Accurate constraints on the missing baryons are valuable to improve our understanding of the fundamental parameters that govern the Universe. 

To this aim, several techniques have been adopted. For instance, \citet{Nicastro2018} and~\citet{Nevalainen19} used quasars to search for absorption features that correspond to baryons in the form of neutral gas and the intracluster medium. The dispersion measure of fast radio bursts (FRBs) provides an integrated measurement of baryon density ($\Omega_{\rm b}h^{2}$) out to localized host galaxies at high redshifts~\citep{Macquart2020,Yang22}. Quantified deuterium abundance in the high-redshift metal-poor damped Ly$\alpha$ system (DLAs) can reveal the primordial deuterium abundance, which leads to precise measurements of baryon density~\citep{Cooke18}.

Other probes include searching for baryons using metal absorption lines~\citep{Oh2002,Dave2007,Narayanan2009,Bertone+10,Shull,Keating2014,Tie2022}, statistical stacks on filamentary structures using the thermal Sunyaev\textendash Zeldovich (tSZ) effect~\citep{Tanimura2019,Tanimura20} and X-ray \citep{Tanimura20, Tanimura_Xray22} measurements, and direct X-ray detections of individual filaments \citep{Werner2008, Eckert2015}. Similarly, X-ray diffuse emission~\citep{Galeazi09,Takei11,Fujita2017,Nicola2022} was also used to search for baryons because it is particularly sensitive to the presence of highly ionized plasma with temperatures of $10^{7}\,{\rm K}<T\leq 10^{8}\,{\rm K}$.

The tSZ effect~\citep{Sunyaev1972} is produced by the inverse Compton scattering of the cosmic microwave background (CMB) photons by high-energy electrons. It is sensitive to the properties of the hot gas in galaxy clusters, including its density and temperature. Therefore, by measuring the tSZ effect, we can study the distribution and physical properties of the ionized baryons, which are important for understanding the large-scale structure (LSS) of the Universe and the processes of cluster formation and evolution. The effect imprints a distortion in the CMB blackbody spectrum, which can be expressed as
\begin{eqnarray}
 \frac{\Delta T}{T_{0}}=y S_{\rm SZ}(x) \label{eq:deltaT-T},
\end{eqnarray}
where $S_{\rm SZ}(x)=x\coth(x/2)-4$ is the spectral distortion function, $x\equiv h \nu/{ k_{\rm B}T_{0}}$, and  $T_0=2.725\,{\rm K}$ is the background CMB temperature. The Compton-$y$ parameter in Equation~(\ref{eq:deltaT-T}) is the integral of the electron pressure along the line of sight:
\begin{eqnarray}
y =\frac{k_{\rm B} \sigma_{\rm T}}{m_{\rm e}c^{2}} \int n_{\rm e}T_{\rm e} \diff l\,,
\label{eq:y}
\end{eqnarray}
where $m_{\rm e}, T_{\rm e}$, and $n_{\rm e}$ are the mass, temperature, and number density of electrons, respectively. $\sigma_{\rm T}, k_{\rm B}$, and $c$ are the Thomson scattering cross section, the Boltzmann constant, and the speed of light, respectively. Unlike X-ray luminosity, the tSZ signal is linearly proportional to the baryon density and the electron temperature, making the detection of gas with low densities and medium temperatures possible. To achieve this aim, stacking analyses have been used at the locations of clusters and superstructures to achieve the goal of detecting such gas. Recently,~\citet{Tanimura2019} stacked $\sim 10^{5}$ pairs of luminous red galaxies on the Planck tSZ map and obtained the first detection of warm gas along the filamentary structures. A complementary study of the gas density and temperature was performed in \citet{Tanimura20}, by stacking 24,544 filaments (across $18$\textendash $30\,{\rm Mpc}$ scales) identified by the DisPerSE method~\citep{Sousbie11}. 
\citet{de-Graaff2019} conducted a similar study with CMASS galaxies at higher redshifts and obtained results consistent with the predicted WHIM properties. Other relevant studies can be found in~\citet{Mittaz1998}, \citet{Eckert2015}, \citet{Vavagiakis2021}, \citet{Kusiak2021} and~\citet{Bonamente2022}.

Similarly, the tSZ auto-angular power spectrum and cross-correlation with other LSS tracers have been extensively studied to trace gas distribution, calibrate cluster masses, and characterize cosmic structures on large scales. This is due to the particular advantage of the tSZ effect of not being very sensitive to redshift evolution, which leaves the possibility of cross-correlating with other LSS tracers to pull out the signal of interest. For example, \citet{V.Waerbeke2014} cross-correlated the tSZ map with the Canada\textendash France\textendash Hawaii weak lensing data and obtained a $\sim 6\sigma$ confidence level (CL) detection. \citet{Ma2015} and \citet{Hojjati2017} found that the detected signal in~\citet{V.Waerbeke2014} corresponds to $\sim 50\%$ of the baryons lying beyond the virial radius of the halos with a temperature of $7\times 10^{5}\,{\rm K}\leq T  \leq 3 \times 10^{8}\,{\rm K}$. \citet{Hurier2018} combined the tSZ effect, X-rays, and weak-lensing auto- and cross-correlation angular power spectra to estimate the cluster mass bias and obtained a result consistent with the Planck measurement at $2\sigma$ CL. \citet{Ma2021} used the tSZ-lensing cross-correlation to constrain the pressure profile of galaxy clusters. A series of other studies based on cross-correlating the tSZ results with other tracers to measure the cluster mass bias can be found in~\citet{Ma2017}, \citet{Bolliet18}, \citet{Li2018}, \citet{Makiya2018}, \citet{Salvati-bias}, \citet{Nick2020} and~\citet{Ibitoye22}. 

The cross-correlation technique equally allows us to study the statistical relationship between LSS tracers and other cosmological probes, providing valuable insights into the underlying cosmological parameters such as the Hubble parameter ($h$), matter density ($\Omega_{\rm m}$), and the amplitude of matter density fluctuations ($\sigma_{8}$). For example, \citet{Ken-Osato2019} cross-correlated the tSZ maps from Planck with the Subaru Hyper Suprime-Cam (HSC) year one data and achieved the constraint of $\Omega_{\rm m}=0.3149\pm0.008$ and $\sigma_{8}=0.8304\pm0.014$ (68\% CL.). \citet{KIDS_1000XtSZ} used the tSZ map from Planck and Kilo-Degree Survey (KiDS) data to the same aim, constraining $\Omega_{\rm m}=0.342^{+0.042}_{-0.037}$ and $\sigma_{8}=0.751^{+0.02}_{-0.017}$ (68\% CL.). Similar studies can be found in \citet{DESY1_cosmic_shear}, \citet{HSCY1_CosmicShear}, \citet{HSCY1_CLs}, \citet{Planck2018TT}, \citet{DESY3_galGAl} and \citet{DESY3_CosmicShear}.


Our study pursues a similar scientific goal. We constrain the cosmological parameters including the Hubble parameter, matter density, and the amplitude of matter density fluctuations, together with the astrophysical parameter $\widetilde{W}^{\rm SZ}$ (the product of the mean electron density $\bar{n}_{\rm e}$, electron temperature $T_{\rm e}$, gas bias $b_{\rm gas}$ at redshift $z=0$). To achieve this, we cross-correlate the tSZ effect with the integrated Sachs\textendash Wolfe (ISW) effect (\citealt{ISW}), which characterizes the largest-scale perturbations in the Universe, and use this to provide valuable constraints on WHIM properties and the underlying cosmology. The ISW effect is a secondary anisotropy of the CMB, which arises when a CMB photon from the last scattering surface enters and leaves the time-evolving gravitational potential, which can change its net energy. Therefore, it causes an additional temperature anisotropy on large scales and is sensitive to the growth rate of cosmic structures. Similarly, because the tSZ effect is also sensitive to the growth rate of cosmic structures, the cross-correlation of the ISW with the tSZ effect can provide useful insight into the gas properties on very large scales~\citep{Creque-Sarbinowski16}.

We will employ the ISW and tSZ maps obtained with the Planck satellite for this analysis. The Planck Collaboration detected the ISW signal, with significances from $\sim 2.5 \sigma$ to $\sim 4 \sigma$ CL, by cross-correlating the Planck CMB map with radio sources from the NVSS catalog, galaxies from the optical Sloan Digital Sky Survey (SDSS) and the Wide-field Infrared Survey Explorer (WISE) surveys, and the Planck 2015 convergence lensing map~\citep{Planck2013_XIX, Planck_ISW_2016}. In addition, the Planck Collaboration also provided different maps of the ISW fluctuations, from different combinations of the abovementioned LSS tracers. We will consider the potential systematics from the cosmic infrared background (CIB) in the cross-correlation, which can contaminate both ISW and tSZ maps at $z<1$. We aim to conduct a joint analysis of both the ISW and tSZ effects and constrain the aforementioned cosmological parameters and gas quantities on very large scales.

This paper is arranged as follows. In Sec.~\ref{sec:data}, we detail the observables, including the theoretical models and data sets employed. In Sec.\ref{sec:cross-correlation_detection}, we present the detailed cross-correlation analysis and demonstrate the significance of the cross-correlated signal. In Sec.~\ref{sec:PowerSpectrum}, we carry out the power spectrum analysis. In Sec.~\ref{sec:param_estimation}, we describe the parameter constraints and their cosmological inference. We present our conclusions in Sec.~\ref{sec:conclusions}. Throughout the paper, we assume a spatially flat $\Lambda$CDM model in which, apart from the parameters being varied in Table~\ref{tab:estimates}, we fix the other cosmological parameters to be $n_{\rm s}=0.965$, $\tau=0.0540$, and $\ln(10^{10}A_{\rm s})=3.043$~(\citealt{Planck2018TT}).
\section{Data and Measurements}
\label{sec:data}

\begin{figure}
	\centerline{\includegraphics[width=3.3in]{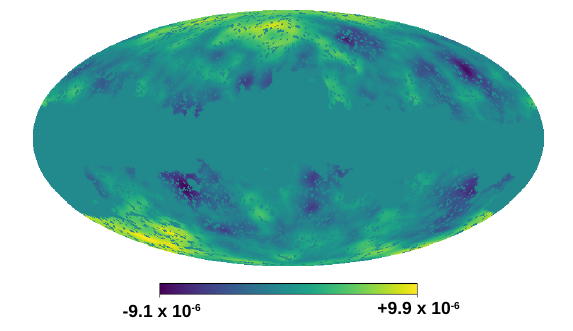}}
	\caption{The ISW map ($\Delta T/T_{0}$) with a 40\% Galactic-plane mask superimposed.}
	\label{fig:ISW_map and mask}
\end{figure}

\begin{figure}
    \centering
	\includegraphics[width=3.3in]{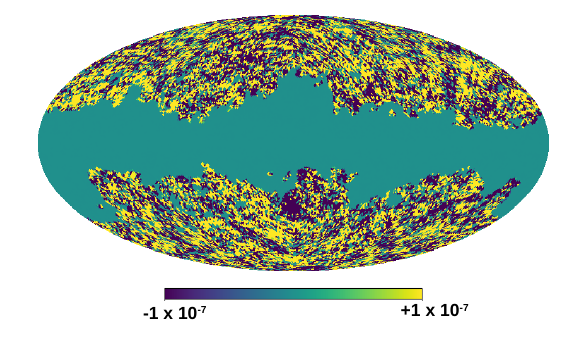}
	\caption{All-sky Compton parameter map ($y$-map) measured by Planck based on the NILC algorithm. The point-source mask and the 40\% Galactic-plane mask have been applied.}
	\label{fig:y-map}
\end{figure}

\begin{figure*}
	\centering

		\centerline{\includegraphics[width=3.8in]{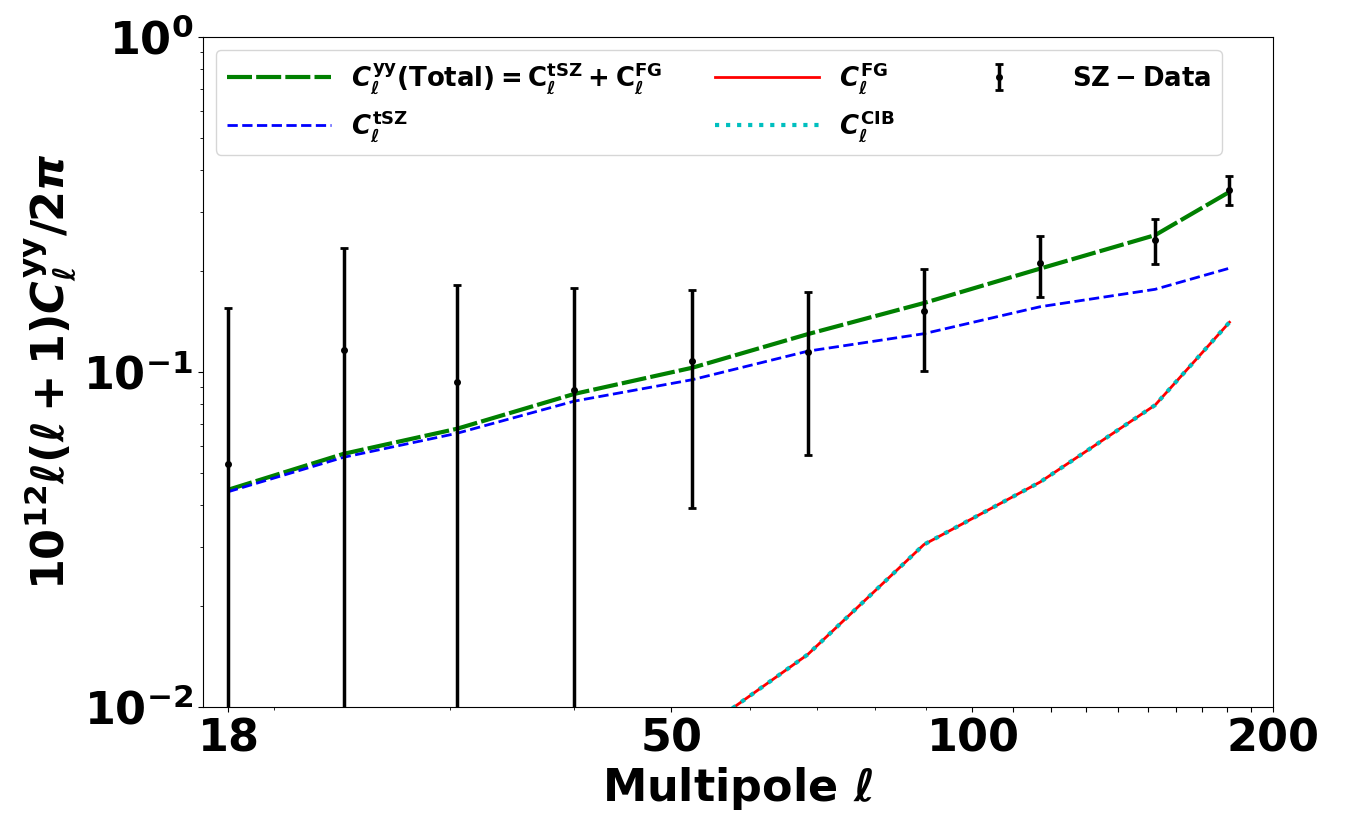}
			\includegraphics[width=3.8in]{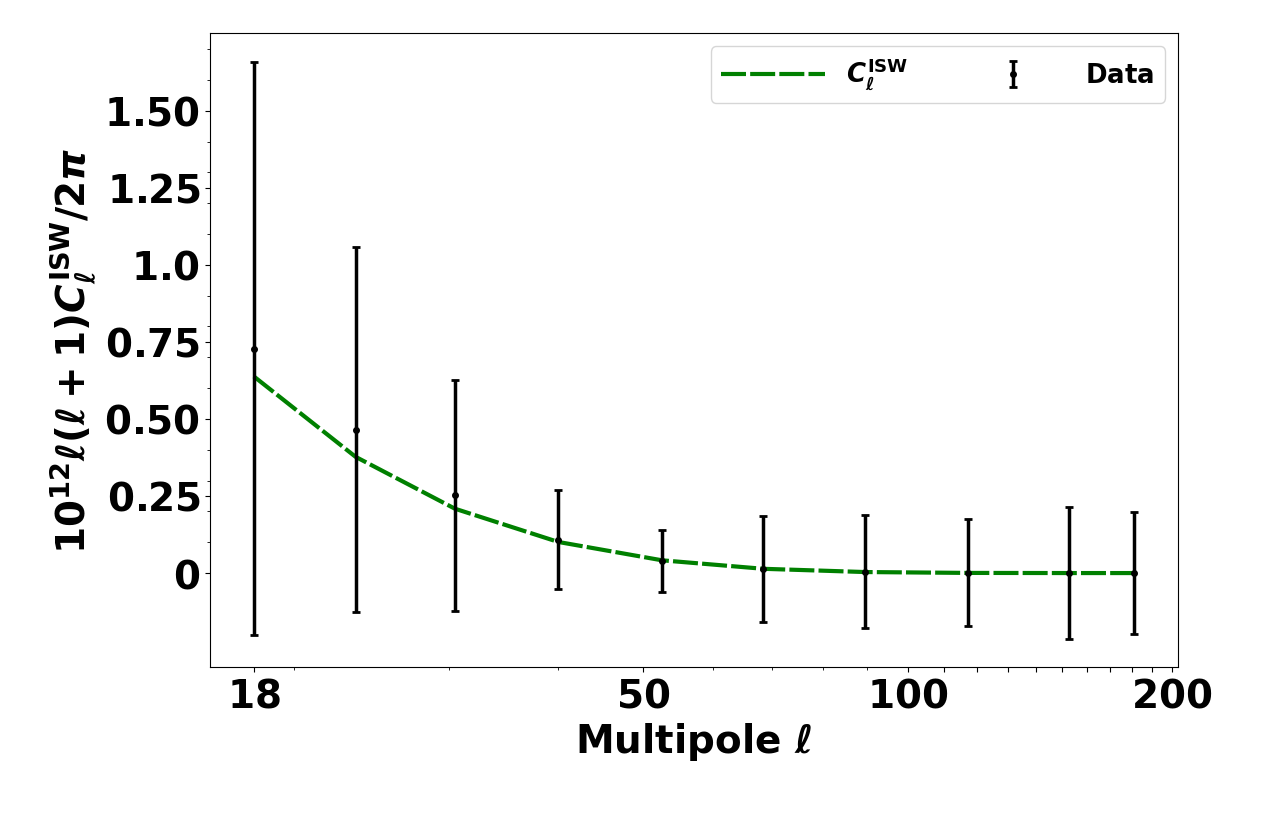}}
		\centerline{\includegraphics[width=12cm]{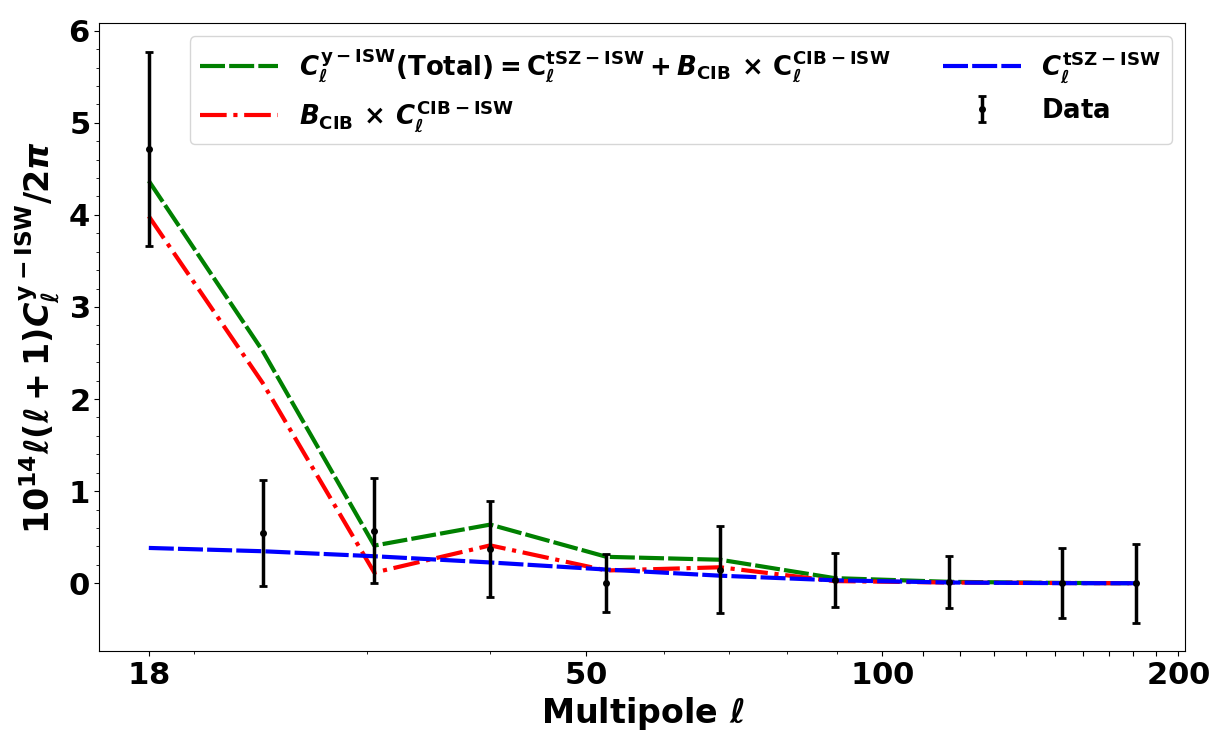}}
	\caption{The measurements of the power spectra, $C^{yy}_{\ell}$ ($y$-auto, {\it upper-left}), $C^{\rm TT}_{\ell}$ (ISW-auto, {\it upper-right}) and $C^{y{\rm T}}_{\ell}$ (ISW-$y$ cross, {\it lower}). Black dots and bars in each case are the means of the spectrum measured from the associated map and the error bars are estimated as the square root of the diagonal values of the corresponding covariance matrix. Other curves have been defined in the legend of each plot.}
	\label{fig:spectra}
\end{figure*} 

\begin{table*}
\centering{
\large
\caption{Summary of Power Spectra Measurements Averaged over [$\ell_{\rm min}$, $\ell_{\rm max}$] Multipole Intervals. Note. For each effective multipole $\ell_{\rm eff}$, we report the values of the power spectra labeled as $C^{yy}_{\ell}, C^{\rm TT}_{\ell}, C^{{\rm T}y}_{\ell}$. We also report the Gaussian 
	error contribution estimated using the \textsc{Namaster}-based covariance, the non-Gaussian error contribution estimated using the \textsc{pyccl}-based covariance
 and the simulation-based covariance described in Sections~\ref{ssec:Cov_matrix} and \ref{ssec:simulation}, labeled as $\sigma_{\rm G}(C^{yy}_{\ell}), \sigma_{\rm NG}(C^{yy}_{\ell}),  \sigma_{\rm TT}$, and $\sigma_{{\rm T}y}$, respectively.}
\label{tab:spectra}	
\renewcommand{\arraystretch}{1.25}

\begin{tabular}{lccccccccc}
\hline 
  $\ell_{\rm min}$ & $\ell_{\rm max}$ & $\ell_{\rm eff}$  & $10^{16} C^{yy}_{\ell}$ & $10^{16}\sigma_{\rm G}(C^{yy}_{\ell})$ & $10^{15} \sigma_{\rm NG}(C^{yy}_{\ell})$&  $ 10^{15} C^{\rm TT}_{\ell}$  & $  10^{14}\sigma^{\rm TT}_{\ell}$ &  $ 10^{16} C^{{\rm T}y}_{\ell}$ &$10^{16} \sigma^{{\rm T}y}_{\ell}$ \\

\hline
		16   & 21   & 18.0   & 9.72077  & 4.56072 & 1.81210 & 13.3760  & 1.70741   & 8.66413  & 1.93446  \\
		21   & 27   & 23.5   & 12.6576  & 2.74956 & 1.01802 & 5.08154  & 0.64732   & 0.59105  & 0.62952  \\
		27   & 35   & 30.5   & 6.08801  & 1.37876 & 0.55815 & 1.65065  & 0.24557   & 0.37412  & 0.37413  \\
		35   & 46   & 40.0   & 3.42842  & 0.62237 & 0.29008 & 0.41536  & 0.06169   & 0.14142  & 0.20050  \\
		46   & 60   & 52.5   & 2.42630  & 0.34870 & 0.14811 & 0.08813  & 0.02255   & 0.00072  & 0.07088  \\
		60   & 78   & 68.5   & 1.51277  & 0.16553 & 0.07488 & 0.01740  & 0.02262   & 0.01933  & 0.06264  \\
		78   & 102  & 89.5   & 1.17419  & 0.07292 & 0.03873 & 0.00380  & 0.01425   & 0.00285  & 0.02281  \\ 
		102  & 133  & 117.0  & 0.96084  & 0.03204 & 0.01964 & 0.00071  & 0.00795   & 0.00063  & 0.01297  \\
		133  & 173  & 152.5  & 0.66389  & 0.01251 & 0.00999 & 0.00010  & 0.00572   & 0.00007  & 0.01006  \\
		173  & 190  & 181.0  & 0.66705  & 0.00993 & 0.00642 & 0.00001  & 0.00377   & 0.00001  & 0.00815  \\ 
\hline

\end{tabular}
}
\end{table*}

\begin{figure*}
	\centering
	\includegraphics[width=17cm]{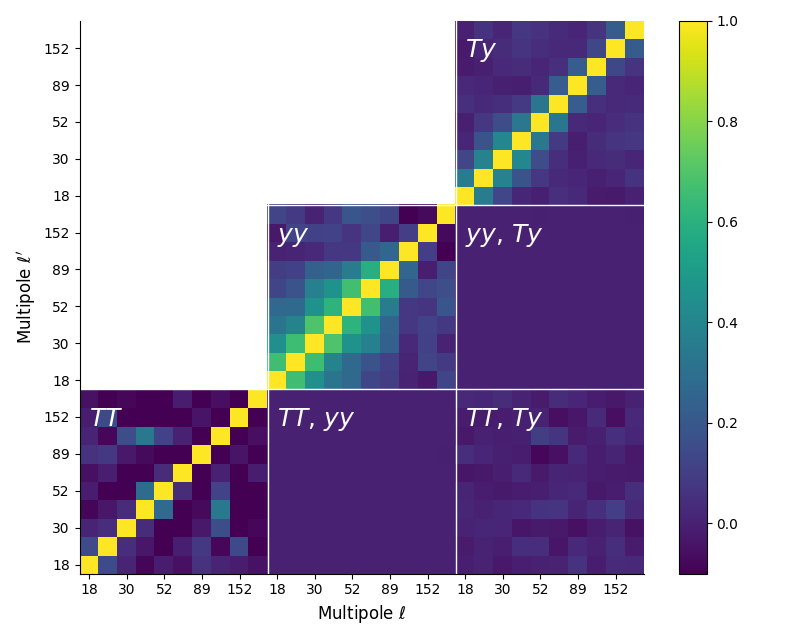}
	\caption{The six independent blocks of the full correlation matrix defined in Equation~(\ref{eq:correlation}). The diagonal blocks show the standard correlations for the TT, $yy$, and T$y$ spectra; the offdiagonal blocks show the additional cross-correlations between different spectra. The correlation values are shown for pairs of the effective band power multipoles defined in Table~\ref{tab:spectra}.}
	\label{fig:correlation}
\end{figure*}


\subsection{ISW Map}
\label{ssec:ISW-Data}
The ISW effect~\citep{ISW} is a secondary anisotropy of the CMB caused by the time-varying gravitational potentials of the LSS, which dominates the CMB on large scales. At low redshifts, it shows a direct signature of dark energy in a $\Lambda$CDM Universe because of the potential decay effect. ISW is also sensitive to the evolution of the growth factor, so it can test alternative theories of gravity along with other cosmological probes. However, because the ISW signal is mostly on large scales that are dominated by cosmic variance, the number of modes extractable is limited~\citep{Maniyar2019}. Thus, the ISW effect is only detectable by cross-correlating with other LSS observations~\citep{Crittenden1996,Planck_ISW_2016}. Figure\,\ref{fig:ISW_map and mask} is the Planck's ISW map delivered in \textsc{HEALPix}\footnote{The \textsc{HEALPix} software has been extensively used to handle pixelated data on the sphere, such as~\citet{Tanoglidis2021}, \citet{Appleby2021}, \citet{Saydjari2022}, and \citet{Chen2022}.} format~\citep{Gorski2005} with an original pixel resolution $N_{\rm side}=64$. We notice that the resolution of the ISW map is $160^{\prime}$ in full width at half maximum (FWHM). As indicated in \citet{Planck_ISW_2016}, this map has a negligible contribution from foreground residuals.

\subsection{Compton parameter map} 
\label{ssec:tSZ-Data} 
In this analysis, we employ the full-sky Compton-$y$ map provided by the Planck satellite mission (Figure~\ref{fig:y-map};~\citealt{Planck2015_tSZ_paper}).
The Compton-$y$ map is the result of the combination of  Planck individual frequency maps convolved to a resolution of $10^{\prime}$ using either the Modified Internal Linear Combination Algorithm (MILCA) method~\citep{hurier13} or the Needlet Independent Linear Combination (NILC) algorithm~\citep{remazeilles11}. Because these two maps are consistent in the limit of uncertainties, for the rest of this work we only present the results for the NILC map. To block out spurious contamination and Galactic foregrounds that can affect our results, we combine the 40\% Galactic mask with the point-source masks. The Compton parameter map, Galactic mask, and point-source masks are publicly available on the Planck Legacy Archive\footnote{http://pla.esac.esa.int/pla}.

To match the resolution of the ISW map, we deconvolve the Compton-$y$ map with a $10^{\prime}$ Gaussian beam and then convolve with a $160^{\prime}$ Gaussian beam to make it the same resolution with the ISW.\footnote{This step is achieved by multiplying $a^{\rm tSZ}_{\ell m}$ by the ratio $B^{160}_{\ell}/B^{10}_{\ell}$ function, where $B^{160}_{\ell}$ ($B^{10}_{\ell}$) is the Fourier transform of a Gaussian beam with a 160 ($10^{\prime}$) FWHM.} We then downgrade the Compton-$y$ map to a coarser pixelization of $N_{\rm side}= 64$, the same as the ISW map. We will use this reprocessed $y$-map in all subsequent analyses.

\subsection{Measurement}
\label{sec:measurement}
We now describe how we obtained our auto- and cross-correlation measurements. 
The Planck ISW and Compton $y$-parameter maps described in Sections.~\ref{ssec:ISW-Data} and Sec.~\ref{ssec:tSZ-Data} are used to measure the projected autocorrelation angular power
spectrum of each observable and the cross-correlation
angular power spectrum of the two. This analysis is conducted using the publicly available pseudo-$C_{\ell}$ MASTER algorithm~\citep{Hivon02}, as implemented in the \texttt{NaMaster} package\footnote{The code uses the MASTER (pseudo-$C_{\ell}$) approach to analyze and compute in full sky the angular auto- and cross-power spectra of masked fields of any pairs of spin for an arbitrary number of known contaminants. The source is hosted at \url{https://github.com/
LSSTDESC/NaMaster} and documented at
\url{http://namaster.readthedocs.io/.}}~\citep{NaMaster}. The code permits dealing with spin 0 and 2 signals, e.g., the case of the CMB temperature and polarization. However, the case considered in this work always deals with scalar quantities. The software is efficient in handling complex masks and high-resolution pixelization. It also executes $E$- or $B$-mode purification in both flat-sky approximation and in full-sky mode. We then made a sanity check with the \textsc{anafast} subroutine included in the \textsc{HEALPix} software package \citep{healpix,Zonca2019}. The power spectra measured using both software packages are in good agreement.

By construction, \texttt{NaMaster} accepts all sky maps in the form of \textsc{HEALPix} maps exclusively with RING ordering and produces the $C^{T_{1}T_{2}}_{\ell}$ power spectrum if two spin$-0$ fields are given, $C^{T_{1}E_{2}}_{\ell}$, $C^{T_{1}B_{2}}_{\ell}$ for one spin$-0$ field and one spin\,$>0$ field, and $C^{E_{1}E_{2}}_{\ell}$, $C^{E_{1}B_{2}}_{\ell}$, $C^{E_{2}B_{1}}_{\ell}$, $C^{B_{1}B_{2}}_{\ell}$ power spectra for two given spin\,$>0$ fields (the subscripts ``1'' and ``2'' represent the field or the observable of interest). Hence, since both the ISW and Compton parameter maps are given as spin$-0$ fields, the measured auto- and cross-angular power spectra\footnote{These power spectra have been corrected for the mode-coupling effect as a result of the survey mask. However, for a check, we also measured the cut-sky power spectra using \texttt{NaMaster} and then manually transformed the spectra to full sky. The results are consistent.} are $C^{\rm TT}_{\ell}$, $C^{yy}_{\ell}$ and $C^{{\rm T}y}_{\ell}$ on full sky respectively. To compare them with the theoretical prediction, we show in Figure~\ref{fig:spectra} the resulting $y$-auto ($yy$), ISW-auto (TT), and ISW-$y$ cross-power spectra (${\rm T}y$).

The sensitive $\ell$-range in this study is $\ell \in [16,190]$. The lower cut is due to the ISW map not having modes for $\ell<8$~\citep{ISW} and the significant cosmic variance between $\ell \in [8,16]$, which can bias our result. The high-$\ell$ cut is due to the limited resolution of the ISW map ($N_{\rm side}=64$). 
The data points we show in Figure~\ref{fig:spectra} are the binned-{$\ell$} ones, to reduce the empirical scatter between adjacent multipole amplitudes $\ell$. We adopt the same scheme for binning as in~\citet{Planck2016}. In addition, this makes the power spectrum coupling matrix invertible to some extent. One may notice that the \citet{Planck2016} used 19 multipole bins in total, but we used 10 bins for the range of $\ell_{\rm eff}=18.0$ to $\ell_{\rm eff}=181.0$ in total. The binning is done by evaluating each bin's average power spectrum values. The resulting values are quoted in Table~\ref{tab:spectra}.

We remind the reader that the $y$-auto power spectrum presented in the upper left of Figure~\ref{fig:spectra} was measured in full sky while the one presented on the left of Figure 11 in~\citet{Planck2015_tSZ_paper} was measured in partial sky. As a comparison, we show in Figure~\ref{fig:Compare_yy_with_planck} in Appendix~\ref{App:Appendix C} the $C^{yy}_{\ell}$ both from our analysis and Planck's partial-sky result. A closer look at Figure 11 in~\citet{Planck2015_tSZ_paper} shows that the $C^{yy}_{\ell}$ measurement is greatly affected by Galactic emissions below $\ell < 20$, but not so much for $\ell \gg 20$. Therefore, it is not surprising to see the consistency between our study and the Planck results for $18.0 < \ell < 152.5$ (see Appendix~\ref{App:Appendix C} for more details).

\subsection{Covariance matrix calculation}
\label{ssec:Cov_matrix}
An accurate estimate of the Gaussian covariance can be calculated using \texttt{NaMaster}. We then generated 1000 realizations of the Compton-$y$ map from the fiducial power spectrum. Each of these maps is supplied as a spin\textendash 0 field to the code with the same weighting/masking scheme, to output binned power spectra $C^{i}_{\ell}$ (i.e. $\hat{C}^{yy, i}_{\ell}$); the resultant covariance can then be calculated from the binned power spectra. 

This covariance matrix estimated from \texttt{NaMaster} is Gaussian and does not account for correlation between different multipoles. However, hydrodynamical simulations showed that SZ fluctuations can indeed be non-Gaussian~\citep{Seljak2001,Zhang_pen_Wang_2002}. Therefore, to obtain accurate constraints on cosmological parameters, it is important to include the non-Gaussian nature of the SZ fluctuations. To this aim, we calculated the connected non-Gaussian covariance matrix using the \texttt{pyccl}\footnote{The Core Cosmology Library (CCL) is an open-source software package written in the C programming language, with a Python interface, which is publicly available at \url {https://github.com/LSSTDESC/CCL}. It computes to a high degree of accuracy several cosmological quantities, which include but are not limited to angular power spectra, correlation functions, halo bias, and the halo mass.} code~\citep{pyccl_code19} to capture the coupling between the measurements at multipoles $\ell$ and $\ell^{\prime}$. This makes the offdiagonal elements nonzero, by an amount that depends on the parallel configurations of the connected
trispectrum. The resultant error bar plotted in the upper left panel of Figure~\ref{fig:spectra} is then calculated from the sum of the Gaussian and non-Gaussian covariance matrix in the SZ solution.

On the other hand, to accurately account for the statistical information encoded in the ISW sky map, and its cross-correlation with the Compton map, we used Monte Carlo (MC) simulations to generate a set of $N_{\rm sim}$ ISW sky maps (where $N_{\rm sim}= 900$), as described in Section~\ref{ssec:detection}. From each of these simulated maps, we output the binned power spectra as $\hat{C}^{{\rm TT}, i}_{\ell}$ and their respective cross-correlation power spectra with the tSZ map as  $\hat{C}^{{{\rm T}y}, i}_{\ell}$. We then compute the covariance matrices using

\begin{eqnarray}
 \label{eq:NM_cov}
{\rm Cov^{\tt NM}}(\hat{C}^{\rm A}_{\ell_{\rm eff}}, \hat{C}^{\rm B}_{\ell^{\prime}_{\rm eff}}) &=& \frac{1}{N_{\rm band}} \sum^{N_{\rm band}}_{i=1} (\hat{C}^{{\rm A},i}_{\ell_{\rm eff}} - \bar{C}^{\rm A}_{\ell_{\rm eff}}) \nonumber \\
& \times & (\hat{C}^{{\rm B},i}_{\ell^{\prime}_{\rm eff}} - \bar{C}^{\rm B}_{\ell^{\prime}_{\rm eff}}).
\end{eqnarray}
The average of the resampled power spectra is given by 
\begin{eqnarray}
 \bar{C}^{\rm A}_{\ell} = \frac{1}{N_{\rm band}} \sum^{N_{\rm band}}_{i=1}\hat{C}^{{\rm A}, i}_{\ell_{\rm eff}},
\end{eqnarray} 
where ${\rm A}$ stands for either ${\rm TT},yy$, or ${\rm T}y$, and $N_{\rm band}=10$ is the number of bins in $\ell$-space. 

To maximize and optimize the information encoded in the measurements, we concatenate all three measured power spectra into a vector: 
\begin{eqnarray}
\mathbf{C}_{\ell}^{\rm Tot}\equiv \left(C^{\rm TT}_{\ell}, C^{{\rm T}y}_{\ell}, C^{yy}_{\ell} \right). \label{eq:Cell_tot}    
\end{eqnarray}
Then the total covariance matrix is a $3\times 3$ block matrix, and each block is an $N_{\rm band}\times N_{\rm band}$ matrix, as
\begin{align}
\label{eq:covfull}
{\rm Cov}^{\rm Total} &= \left\langle \mathbf{C}^{\text{Tot}}_{\ell}{\mathbf{C}^{\text{Tot}}_{\ell}}^{\rm T} \right\rangle \nonumber \\
&= \left[
\begin{array}{ccc}
{\rm Cov}^{\rm TT,TT} & {\rm Cov}^{{\rm TT,T}y} & {\rm Cov}^{{\rm TT},yy} \\
\left({\rm Cov}^{{\rm TT,T}y} \right)^{\rm T} & {\rm Cov}^{{\rm T}y, {\rm T}y} & {\rm Cov}^{{\rm T}y, yy} \\
\left({\rm Cov}^{{\rm TT},yy} \right)^{\rm T} & \left({\rm Cov}^{{\rm T}y, yy} \right)^{\rm T} & {\rm Cov}^{yy,yy} \\
\end{array}
\right],
\end{align}             
where, for instance, the ${\rm Cov}^{{\rm TT,T}y}$ is the cross-covariance obtained using $\hat{C}^{{\rm TT},i}_{\ell}$ and $\hat{C}^{{{\rm T}y},i}_{\ell}$ in Equation~(\ref{eq:NM_cov}).

To evaluate the correlation between different blocks of the covariance matrix, we calculate the correlation coefficient matrix with Equation~(\ref{eq:covfull}) using the expression
\begin{eqnarray}
{\rm Corr}^{\rm A,B}(\ell,\ell^\prime)=\frac{{\rm Cov}^{\rm A,B}(\ell,\ell^\prime)}{\left ({\rm Cov}^{\rm A,A}(\ell,\ell^\prime)\right)^{\frac{1}{2}} \left ({\rm Cov}^{\rm B,B}(\ell,\ell^\prime)\right)^{\frac{1}{2}}}\,.
\label{eq:correlation}
\end{eqnarray}
We show the correlation matrix in Figure~\ref{fig:correlation}. One can see that the offdiagonal matrix blocks, i.e. (TT,$yy$), (T$y$,$yy$), and (TT,T$y$) have much smaller values than the diagonal blocks, indicating a small correlation between the power spectra $C^{y{\rm T}}_{\ell}$, $C^{yy}_{\ell}$, and $C^{\rm TT}_{\ell}$. 

We can also calculate the significance of our observed spectra against the null detection as $S/N \equiv \sqrt{\chi^{2}}$, where 
\begin{eqnarray}
\label{eq:SNR}
\chi ^{2}= \sum_{\ell \ell^{\prime}}^{N} \mathbf{C}^{\rm Tot}_{\ell} \,\,\,{\left({\rm Cov}^{\rm Tot}\right)^{-1}}_{\ell \ell^{\prime}}\,\,\mathbf{C}^{\rm Tot}_{\ell^{\prime}},
\end{eqnarray} and the signal-to-noise ratios (${\rm S/Ns}$) for ${{\rm T}y}$ and $yy$ are $3.6 \sigma$ and $25 \sigma$, respectively, with no significance on the ${\rm TT}$ correlation (See Section~\ref{ssec:detection} for more details). In Sec.~\ref{sec:Hubble_tension}, we see that using $C^{\rm TT}_{\ell}$ only is very weak for constraining cosmological parameters.

\section{Cross-correlation detection} 
\label{sec:cross-correlation_detection}
In this section, we summarize the approach adopted to characterize the tSZ and the ISW by means of a null hypothesis test, by comparing the observed cross-correlation for the data with that obtained from the actual tSZ and a set of simulated ISWs constrained in the same way as the ISW data. This detection approach is similar to the one used in the literature to detect the ISW effect~\cite[e.g.,][] {Boughn2004,Vielva2006}.

\subsection{Simulations}  
\label{ssec:simulation}
The ISW simulations used to characterize the null hypothesis are the ones described in Section 5 of \cite{Planck_ISW_2016}. These simulations contain the statistical properties of the Planck ISW map and are built by applying the linear covariance-based (LCB) filter introduced in \cite{Barreiro2008} and extended in \cite{Manzotti2014} and \citet{Bonamente2022}. They start from coherent galaxy density number simulations of different LSS tracers (the same used to derive the Planck ISW map), which are afterward combined with the LCB to map out the expected ISW anisotropies. Among the different survey combinations described in \cite{Planck_ISW_2016}, we adopt the one containing information from the CMB, and the NVSS, SDSS, WISE, and Planck lensing LSS tracers \cite[see Section 5 in][for details]{Planck_ISW_2016}. We used 1000 simulations generated in this way to characterize the mean value and the full covariance of the tSZ\textendash ISW cross-correlation, as explained in the next subsection.

\subsection{Detection}  
\label{ssec:detection}
\begin{figure*}
\centerline{\includegraphics[width=19cm]{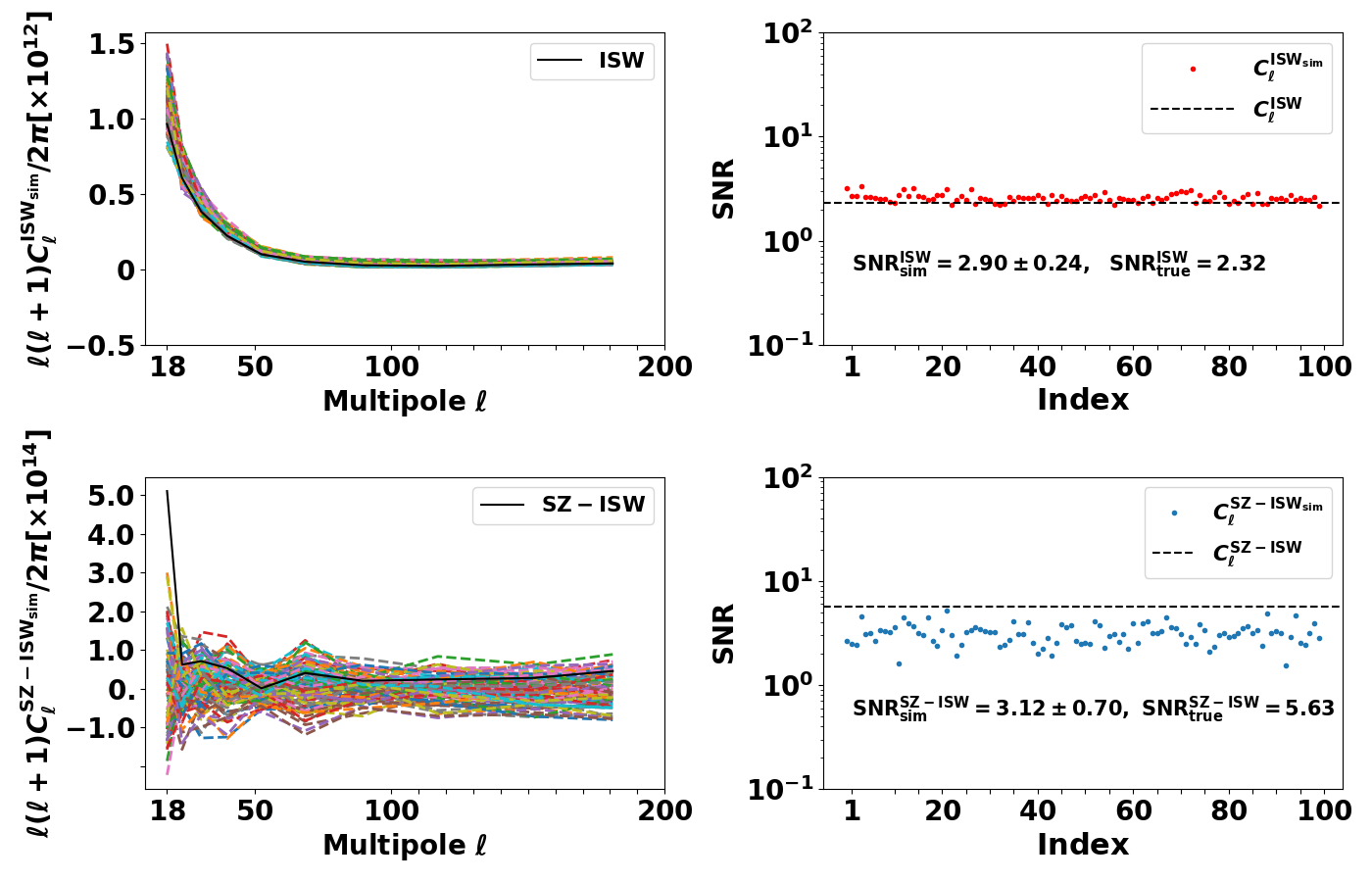}}
	\caption{{\it Upper Left}-the measurements of the ISW auto-power spectrum using the true Planck ISW map (black solid line) and the $100$ simulated ISWs map (colored lines). {\it Upper Right}-the respective ${\rm S/Ns}$ of the $100$ simulated ISW maps (red dots; with mean $2.90$ and rms $0.24$) and the true ISW map (black dashed line; $2.32$). {\it Lower Left}-the cross-correlation power spectra of the $100$ simulated ISWs map with the Planck tSZ map (colored lines) and the true Planck ISW map cross-correlation. {\it Lower Right}-the significance of the true ISW\textendash tSZ cross-correlation (black dashed line; $5.63$) and the $100$ simulated ones (blue dots; with mean $3.12$ and rms $0.70$.).}
	\label{fig:detection}
\end{figure*}
To quantify the statistical significance of our measurement, we perform a robust null hypothesis test from the MC simulations described in \ref{ssec:simulation}. We remark that with our simulation package, we produced 1000 simulated ISW sky maps. Among these, 900 were used to compute the covariance matrix, as detailed in Section~\ref{ssec:Cov_matrix}, while the remaining 100 (name it, $N_{\rm sig}$) were used to quantify the significance as follows. We first measured the spectra from these maps and grouped them into a matrix with size $N_{\rm sig} \times N_{\rm band}$ dimensions. We then computed the $\chi^{2}$ associated with each of these power spectra and their respective ${\rm S/Ns}$~\footnote{We remind the reader that when computing the ${\rm S/Ns}$ associated with each of the 100 spectra, the covariance matrix used in Equation~(\ref{eq:SNR}) is then the one estimated using the 900 independent simulations.} using Equation~(\ref{eq:SNR}) (see Figure~\ref{fig:detection}), from which we estimated their mean and dispersion (rms value). We remark that when such a procedure is followed using the ISW simulations, we are getting a distribution of this statistic (S/R) for the null hypothesis (e.g., there is no correlation). For the cross-power spectrum measurement, the true ${\rm S/N}$ is $5.63\sigma$ (using the ISW sky map obtained from Planck), shown as a black solid line in the lower left panel of Figure~\ref{fig:detection} and as a black dashed line in the lower right panel of Figure~\ref{fig:detection}. The realizations from $100$-simulations are distributed with ${\rm S/N}=3.12\pm 0.7$. Therefore, under the no-correlation hypothesis, the true data deviate from the simulated ones with $(5.63\textendash3.12)/0.7 \simeq 3.6\sigma$. This implies that the no-correlation hypothesis is rejected by $3.6\sigma$ ($\simeq 99.97\%$ CL.), indicating the significance of true ISW\textendash tSZ cross-correlation is achieved at $3.6\sigma$ CL.

\section{Power Spectrum Analysis}
\label{sec:PowerSpectrum}

One can see from Figure~\ref{fig:spectra} that the measurements of the ISW auto-power spectrum and the ISW\textendash tSZ cross-power spectrum are mainly on large scales. This is because the ISW effect primarily exists at low multipoles, due to the late-time decay of the gravitational potential. For this reason, the ISW map from Planck has a low resolution ($\theta_{\rm FWHM}=160^{\prime}$) that smears out the structures beyond $\ell \gtrsim 100$. Therefore, in the following, we will calculate the theoretical power spectrum based on large-scale gas bias models.

\subsection{Compton-$y$ parameter}
\label{sec:Compton-y}
We adopt a large-scale gas bias model for the tSZ effect. Following~\citet{Goldberg99} and~\citet{V.Waerbeke2014}, the gas density contrast is given by $\delta_{\rm gas}(\mathbf{x})=b_{\rm gas}(z)\delta_{\rm m}(\mathbf{x}=\chi\hat{\mathbf{n}},\chi(z))$, where $\delta_{\rm m}$ is the matter density contrast and $b_{\rm gas}(z) \propto (1+z)^{-1}$ is the gas bias. Since we correlate the fluctuation of gas density, we have:
\begin{eqnarray}
n_{\rm e}(\mathbf{x}, z) &= & \overline{n}_{\rm e}(z)\delta_{\rm gas}(\mathbf{x}, z) \nonumber \\
&=& \left[\overline{n}_{\rm e0}\cdot a^{-3}\right] \cdot \left[b_{{\rm gas}, 0}\cdot a\right]\delta_{\rm m}(\mathbf{x},z) \nonumber \\
&=& \overline{n}_{\rm e0}\,b_{{\rm gas}, 0}\,a^{-2}\,\delta_{\rm m}(\mathbf{x},z),
\label{eq:ne}
\end{eqnarray}
where $\bar{n}_{\rm e0}$ and $b_{{\rm gas}, 0}$ are the electron number density and the gas bias at the present day, respectively.

In this model of large-scale gas bias~\citep{Goldberg99,V.Waerbeke2014}, the spatial fluctuation of the gas temperature is ignored and the average temperature is proportional to $a$, i.e., $T_{\rm e}\propto a$. Therefore, for the electron temperature, we have $T_{\rm e}(z)=T_{\rm e}(0)a$. Substituting Equation\,\eqref{eq:ne} into Equation\,\eqref{eq:y}, we have
\begin{eqnarray}
y(\hat{\mathbf{n}}) &=& \left(\frac{\sigma_{\rm T}k_{\rm B}}{m_{\rm e}c^{2}} \right)\int \bar{n}_{\rm e0}b_{{\rm gas}, 0}a^{-2}\delta_{\rm m}\left( \mathbf{x}, z\right)T_{\rm e}(0)a\,a {\rm d}\chi \nonumber \\
&=& W^{\rm SZ} \int \diff \chi \,\delta_{\rm m}(\chi\hat{\mathbf{n}},\chi(z)), \label{eq:y-n}
\end{eqnarray}
where we have substituted the relation between differential proper distance and the comoving distance $dl=ad\chi$. $W^{\rm SZ}$ becomes a constant factor, \begin{eqnarray}
W^{\rm SZ}&=& \left(\frac{k_{\rm B}\sigma_{\rm T}}{m_{\rm e}c^{2}}\right)\,\overline{n}_{\rm e 0}\,b_{{\rm gas}, 0}\,T_{\rm e}(0)\, \nonumber \\
&=& (4.02 \times 10^{-10}\,{\rm Mpc}^{-1}) 
 \times  \widetilde{W}^{\rm SZ},
\label{eq:WSZ_num}
\end{eqnarray}
where 
\begin{eqnarray}
\widetilde{W}^{\rm SZ} = b_{\rm gas, 0} \left(\frac{\bar{n}_{\rm e 0}}{1\,{\rm m}^{-3}} \right)\left(\frac{k_{\rm B}T_{\rm e}(0)}{0.1\,{\rm keV}} \right).
\label{eq:WSZ_dimensionless}
\end{eqnarray}
In the following, we will omit ``0'' in the subscripts for brevity.

\subsection{Temperature fluctuation of ISW effect}
\label{sec:ISW}
The temperature fluctuation caused by this effect is \citep{ISW_tSZ_02,ISW_tSZ_011}:
\begin{eqnarray} 
\label{eq:ISW temp}
\frac{\Delta T (\hat{\mathbf{n}})}{T} = - \frac{2}{c^2} \int \diff z \frac{\partial \phi}{\partial z} (\hat{\mathbf{n}}, z).
\end{eqnarray}
The Poisson equation relates the gravitational potential with the density fluctuation, which can be inverted to calculate the potential value (see, e.g.~\citealt{Creque-Sarbinowski16}):
\begin{eqnarray}
&& \nabla^{2}_{\rm x} \phi = 4 \pi G \bar{\rho}_{\rm m} a^2 \delta_{\rm m}(\chi,\hat{\mathbf{n}}) \nonumber \\
\Rightarrow && \phi(\mathbf{x} ; z)=-\frac{3}{2} \Omega_{\rm m} H_{0}^{2} \frac{D(z)}{a(z)}\left[\nabla_{\mathbf{x}}^{-2} \delta_{\rm m}(\mathbf{x}, z=0)\right],
\end{eqnarray} 
where $\nabla_{{x}}$ is the gradient in comoving coordinates.

\subsection{Auto- and cross-power spectra}
\label{ssec:aps}
The TT angular power spectrum or the temperature correlation represents the mathematical breakdown of the anisotropic temperature map using spherical harmonics analysis.
The tSZ auto-power spectrum is calculated by carrying out the spherical harmonics decomposition on the sky~\citep{Planck2014_tSZ, V.Waerbeke2014, Ma2015,Creque-Sarbinowski16}:
\begin{eqnarray} 
\label{eq:tSZ_auto}
C^{yy}_{\ell}
&=& \left(W^{\rm SZ}\right)^{2}\int  {\rm d} \chi  \frac{1}{\chi^{2}}P_{\rm m}\left(\frac{\ell +1/2}{\chi(z)},z \right)\, \nonumber \\
&=& \int  {\rm d} \chi   \left(\Delta^{\rm SZ}(\chi) \right)^2 P_{\rm m}\left(\frac{\ell +1/2}{\chi(z)},z \right)\,, \label{eq:Cell-tSZ}
\end{eqnarray}
where we have defined, in the second equality, $\Delta^{\rm SZ}(z)\equiv W^{\rm SZ}/\chi$. $P_{\rm m}\left((\ell +1/2)/\chi,z \right)$ is the linear matter power spectrum, which we compute using the {\sc camb} code~\citep{Lewis2000}.

Similarly, we use Limber approximation to derive the ISW auto-power spectrum weighted by the ISW kernel~\citep{Creque-Sarbinowski16}:
\begin{eqnarray}
\label{eq:ISW_auto}
C^{\rm TT}_{\ell} 
&=&\int {\rm d} \chi \left(\Delta^{\rm ISW}_{\ell} \right)^{2}P_{\rm m}\left(\frac{\ell +1/2}{\chi(z)},z \right)\,,
\end{eqnarray}
where
\begin{eqnarray}
\Delta^{\rm ISW}_{\ell}&=&\frac{3\Omega_{\rm m}H^{2}_{0}}{c^{3}\left(\ell +\frac{1}{2} \right)^{2}}\chi(z)H(z)\frac{1}{D(z)}\frac{\rm d}{{\rm d}z} \left(\frac{D(z)}{a(z)} \right)\, \label{eq:tilde-delta}
\end{eqnarray}
is the redshift-dependent kernel of the ISW.\footnote{Our definition of the kernel has an additional $1/D(z)$ factor to Equation~(6) in~\citet{Creque-Sarbinowski16}, because we use the power spectrum at redshift $z$.} The cross-power spectrum of the ISW and tSZ becomes
\begin{eqnarray}
 \label{eq:cl-ISW_y}
C^{\rm tSZ-ISW}_{\ell} &=&  \int {\rm d} \chi \, \Delta^{\rm ISW}_{\ell}(z) \Delta^{\rm SZ}(z)P_{\rm m}\left(\frac{\ell +1/2}{\chi(z)},z\right),\nonumber \\
\end{eqnarray}
where we have used the definition below Equation~(\ref{eq:Cell-tSZ}). We note that both Equations (\ref{eq:cl-ISW_y}) and ~(\ref{eq:tSZ_auto}) cannot be compared directly with the measurements described in Sections~\ref{ssec:ISW-Data} and~\ref{ssec:tSZ-Data},  because they do not represent the exact information from maps until we factor in the foreground contributions in each data set.

\subsection{Modeling of the contaminations}
\label{sec:FG_modelling_and_redshift_distr}
\begin{figure}
	\centering
	\includegraphics[width=3.3in]{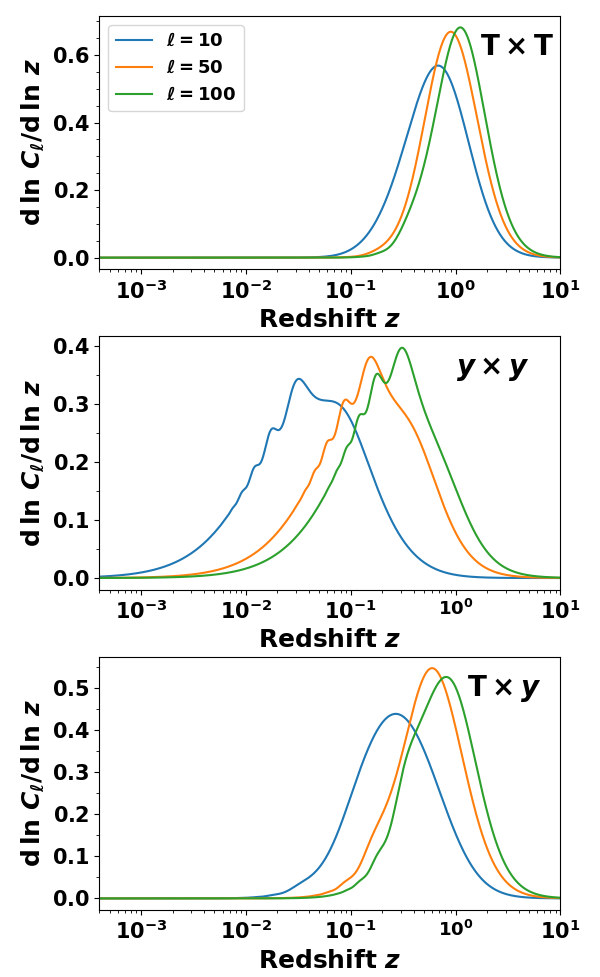}
	\caption{The effective redshift range driving the auto-and cross-power spectra, shown for three reference effective multipoles at $\ell$= 10, 50, and 100.}
	\label{fig:z-distribution}
\end{figure}

The power spectra of the $yy$ and ${{\rm T}y}$ both suffer from contamination. Due to the imperfect cleaning of the Compton-$y$ map by the component separation algorithms, the resulting map contains residual foregrounds from the clustered CIB, emission due to the infrared (IR) point sources, radio point-source emission, and the instrumental correlated noise.\footnote{The instrumental noise is called ``correlated'' because certain $\ell$-dependent features are introduced in the noise due to the NILC algorithm procedure, which makes the noise no longer white.} These foreground components must therefore be accounted for in the $y$ power spectrum calculation. We assume that the residual foregrounds in the map keep the same shapes as their original power spectra but with reduced amplitudes, due to the cleaning process. Therefore, we model the residual foregrounds by scaling them with a set of amplitudes referred to as {\it nuisance parameters}, i.e., $(A_{\rm CIB}, A_{\rm IR}, A_{\rm RS}, A_{\rm CN})$. Then the measured $C^{yy}_{\ell}$ power spectrum can be written as (see also, e.g.~\citealt{Makiya2018,Ibitoye22}):
 \begin{eqnarray}
 \label{eq:clforegrounds}
 C^{yy}_{\ell} &=& C^{\rm tSZ}_{\ell} +A_{\rm CIB}C^{\rm CIB}_{\ell}\nonumber 
 \\
&+& A_{\rm IR}C^{\rm IR}_{\ell}
 + A_{\rm RS}C^{\rm RS}_{\ell}+A_{\rm CN}C^{\rm CN}_{\ell}\;,
 \end{eqnarray}
where $C^{\rm tSZ}_{\ell}$ is the theoretical tSZ effect power spectrum. $C^{\rm CIB}_{\ell}, C^{\rm IR}_{\ell}, C^{\rm RS}_{\ell}$, and $C^{\rm CN}_{\ell}$ are the CIB, IR sources, radio sources, and the correlated noise spectra that may remain in the Compton-$y$ map, which are tabulated in \citet{Bolliet18}. These foregrounds mainly dominate on small scales. For example, the correlated noise power spectrum may only match the $yy$ spectrum amplitude at the multipoles $\ell \simeq 2742$~\citep{Ibitoye22}. In our analysis, because of the resolution of the $y$-map $N_{\rm side}= 64$, the sensible $\ell$-range is up to $\ell_{\rm max} \simeq 192$ (see Figure~\ref{fig:spectra}). Thus, the contribution from the instrumental correlated noise is negligible. Similarly, we neglected
the IR and radio point-source components because they are also subdominant on large angular scales and would only contribute significantly on scales beyond $\ell \sim 200$ (see Table~\ref{tab:spectra} of this paper and also Table~3 in~\citealt{Bolliet18} and Table~3 and Figure~6 in~\citealt{Makiya2018}). Therefore, to recover the observed $yy$ power spectrum, the CIB power spectrum weighted by the $A_{\rm CIB}$ coefficient would be sufficient. Hence, Equation~(\ref{eq:clforegrounds}) is reduced to
\begin{eqnarray}
 \label{eq:clforegroundserrors}
 C^{yy}_{\ell} = C^{\rm tSZ}_{\ell} + A_{\rm CIB}C^{\rm CIB}_{\ell}.
 \end{eqnarray}
 One can see from the upper left panel of Figure~\ref{fig:spectra} that for the low-resolution $y$-map we are using, the CIB contributes the most at $\ell \geq 50$, where the other foregrounds are subdominant. 
 We have also verified that including other foregrounds does not necessarily improve our results (see Figure~\ref{fig:planck_like_and_mine}). For these foregrounds, we use the same power
spectrum templates used in the original Planck analyses, \citet{Planck2014_tSZ} and \citet{Planck_ISW_2016} which is also shown in Table 3 of \citet{Bolliet18}.

We also calculate the redshift dependence of the three cross-correlation power spectra and see what range of redshifts they are most sensitive to. We take the logarithmic derivative of the three power spectra with respect to the redshift:
\begin{eqnarray}
\frac{{\rm d}\ln C^{XY}_{\ell}}{{\rm d}\ln z}
= \left( \frac {\Delta^{XY}_{\ell}(z)}{H(z)}\right)\frac{cz}{C^{XY}_{\ell}} P_{\rm m}\left(\frac{\ell+1/2}{\chi}, z\right),
\label{eq:dCXY-dz}
\end{eqnarray}
where we have substituted ${\Delta^{XY}_{\ell}(z)= \Delta^{X}_{\ell}(z) \Delta^{Y}_{\ell}(z)}$. $X, Y$ can be either the ISW (auto) or SZ (auto) or both (cross). In Equation~(\ref{eq:dCXY-dz}), we fix the model parameters at the best-fitting values in Table~\ref{tab:estimates}. We show the redshift dependence in Figure~\ref{fig:z-distribution}. One can see that although the redshift distributions for the SZ and ISW auto-power spectra are comparable, that of ISW auto-power spectrum falls off at $z \sim 3$. On the other hand, the tSZ\textendash ISW cross-power spectrum has a broad distribution, mainly covering $0.1<z \leq 10$.

\begin{figure}
	\centering
	\includegraphics[width=3.3in]{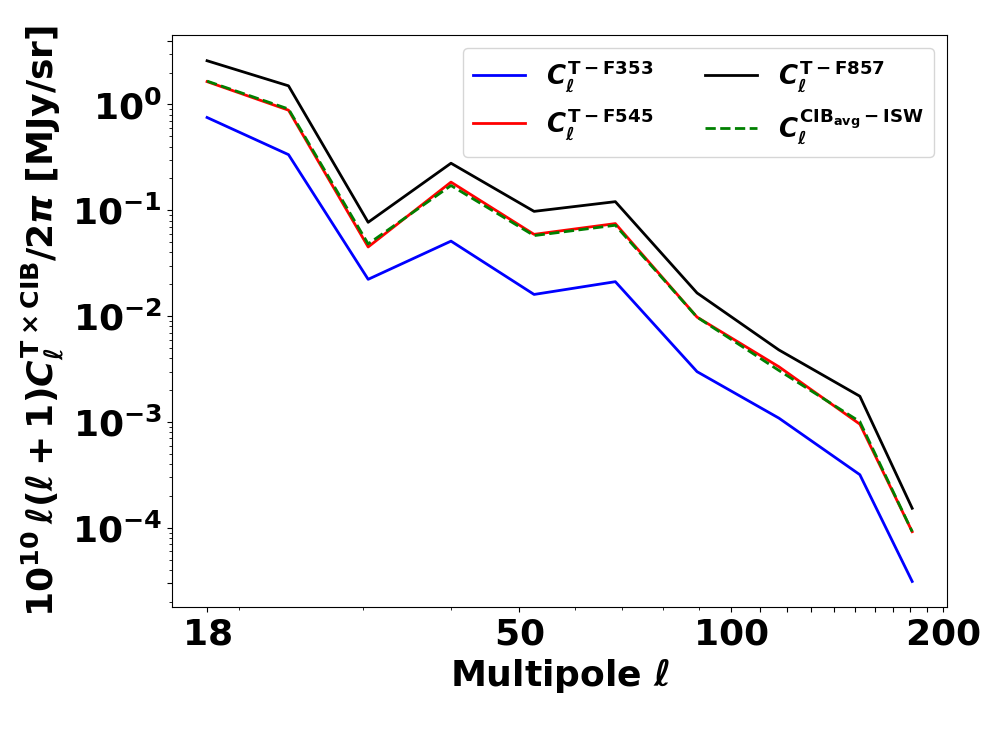}
	\caption{Cross-power spectrum of the CIB in three Planck frequencies with the ISW. The spectrum was measured using the public code \texttt{NaMaster}.}
	\label{fig:T_cross_CIBs}
\end{figure}

Because of the dependence of the ISW on the redshift regime $z \geq 1$, it is necessary to consider the effect of CIB contamination (extragalactic dust) in the $y{\rm T}$ cross-correlation \citep{Makiya2018}. We estimate the level of CIB contamination in the $y{\rm T}$ cross-correlation by using the three individual frequency CMB maps at 353, 545, and 857 GHz from the Planck legacy archive. We degraded the resolution of each CIB map from $N_{\rm side}=2048$ to $N_{\rm side}= 64$ to match the pixelization of the  ISW. We remind the reader that the CIB pixelized maps are at a resolution of $10^{\prime}$. Therefore, we also deconvolve the CIB maps with a $10^{\prime}$ Gaussian beam and then convolve them with a $160^{\prime}$ Gaussian beam to match the ISW map. We then measure the cross-correlation between all three CIB maps and the ISW map and show them in Figure~\ref{fig:T_cross_CIBs}. We find that the cross-correlation of each of the CIB frequency maps with the ISW map is positive, as expected. This is because at $z>1$, where ISW is relevant, the CIB is non-negligible. Similarly, the CMB photons that travel through the time-evolving gravitational potentials that underlie the large-scale cosmic structures are the sources of the CIB \citep{CIB-ISW}. As shown in Figure~\ref{fig:T_cross_CIBs}, the amplitude of the cross-correlation is sensitive to the frequency, which may be because the CIB probes galaxies on low-frequency bands at higher redshifts, while ISW is sensitive to galaxies at low redshifts \citep{Maniyar2019}. Therefore, since the power spectra of the CIB\textendash ISW cross-correlations are very similar in shape but with different amplitudes, we compute the average of the three power spectra, $C^{\rm CIB_{avg}-ISW}_{\ell}$, as the final CIB contamination model. 

The ${\rm T}y$ cross-correlation data are then modeled as 
\begin{eqnarray}
  \label{eq:cl_cib_contamination}
  C^{y-{\rm ISW}}_{\ell} = C^{\rm tSZ-ISW}_{\ell} + B_{\rm CIB}C^{\rm CIB_{avg}-ISW}_{\ell}\,,
\end{eqnarray}
where $C^{\rm tSZ-ISW}_{\ell}$ is the theoretical prediction of the ISW and tSZ cross-power spectrum (Equation~(\ref{eq:cl-ISW_y})). $C^{\rm CIB_{avg}-ISW}_{\ell}$ is the average of the three cross-power spectra of the ISW and CIB. We then introduce a dimensional parameter $B_{\rm CIB}$(with unit ${\rm sr}\cdot {\rm Myr}^{-1}$). This parameter estimates the CIB contamination level in the $y{\rm T}$ cross-correlation analyses. We remind the reader that $B_{\rm CIB}$ is different from but related to the dimensionless $A_{\rm CIB}$. The latter represents the amplitude of the CIB contamination in the $yy$ autocorrelation. The two parameters are related because they model the same foreground contamination in different maps. 

We do not expect to have a significant contribution of galactic foregrounds, because, as indicated in Section \ref{ssec:ISW-Data}, the ISW map is almost free of foreground contamination. Also, for our multipole regime, the NILC tSZ maps seem to be quite robust against different masking scenarios (see Figure 11 in \citealt{Planck2015_tSZ_paper}). In addition, a significant fraction of the ISW signal comes from LSS tracers, which are not correlated with the galactic foregrounds.


\section{Constraints on Parameters}  
\label{sec:param_estimation}
The theoretical prediction modeled in Sec.~\ref{sec:data} allows us to connect the astrophysical, cosmological, and foreground parameters with the observed auto- and cross-power spectra. The cosmological parameters that the power spectra are sensitive to include the fractional matter density ${\Omega_{\rm m}}$, Hubble parameter $h\equiv H_{0}/100\,{\rm km}\,{\rm s}^{-1}\,{\rm Mpc}^{-1}$, and the rms matter density fluctuations at $8\,h^{-1}{\rm Mpc}$, i.e., $\sigma_{8}$. The astrophysical parameter involved is mainly the one related to $\widetilde{W}^{\rm SZ}$, i.e. the product of $b_{\rm gas}$ (gas bias), $\bar{n}_{\rm e}$ (the electron number density today), and $T_{\rm e}$ (the electron temperature at present). This joint parameter quantifies the amplitude of the warm baryonic dark matter on the largest scales. The foreground parameter $A_{\rm CIB}$ gauges the amplitude of the foreground spectrum to the measured $C^{yy}_{\ell}$. $B_{\rm CIB}$ captures the amount of the CIB contamination at the level of the ISW\textendash tSZ cross-correlation. In the following subsection, we describe the details of our methodology to constrain these parameters\footnote{Appendix~\ref{App:consistency} and Figure~\ref{fig:compare_constraints} show the comparison between results from using {\sc emcee} and {\sc ultranest} packages.}.

\begin{table}
	\centering
		\renewcommand{\arraystretch}{1.3}
	\caption{Results of Parameter Estimation. Note. For each parameter, we report the range over which it was fitted with a flat prior (second column) and best-fit values with 68\% CL (third column). All parameters are dimensionless, apart from $B_{\rm CIB}$ in $[\text{sr}\,\text{MJy}^{-1}]$ (We have converted the dimensional constraint on $W^{\rm SZ}$ to the joint constraint on $b_{\rm gas} T_{\rm e}(0)\bar{n}_{\rm e}$). The first three rows on the upper partition are the constraints on the cosmological parameters. The fourth row is the derived constraint on $S_{8}$. The middle partition is the constraint on the gas parameter and the lower one is for foreground parameters. Notice that for the $h$ value, we quote the diagonal block's constraint here (Equation~(\ref{eq:H0_diagonal})) for the reason given in Section~\ref{sec:Hubble_tension}.}
	\label{tab:estimates}		
	\begin{tabular}{lcccc} 		
\hline  
\hline 
Parameter     & Prior (flat) &  68\% CL.   \\
\hline

		$\Omega_{\rm m}$ & $[0.2,0.4]$  &$0.317^{+0.040}_{-0.031}$ \\
		
		$h$ & $[0.6,0.8]$  &$ 0.697^{+0.020}_{-0.015}$\\
		$\sigma_{8}$ & $[0.64,0.9]$ &$0.730^{+0.040}_{-0.037} $\\ 
		
		${S_8= \sigma_{8} ({\Omega_{\rm m}}/0.3)^{0.5}}$ & $[-]$ &$0.755\pm{0.060}$\\  
		\hline

		$b_{\rm gas} \left (T_{\rm e}(0)/0.1\,{\rm keV} \right ) \left ( \bar{n}_{\rm e}/{\rm m}^{-3} \right )$& $[0.02,10]$ &$3.09^{+0.320}_{-0.380}$\\
       \hline
		
		$B_{\rm CIB}\times 10^4$  & $[0,5]$ & $4.0^{+1.200}_{-1.100}$\\ 
		
        $A_{\rm CIB}$ & $[0,5]$ & $0.81^{+0.086}_{-0.079}$\\


		\hline
	\end{tabular}
\end{table}

\subsection{Likelihood method}
\label{ssec:likelihood}
\begin{figure*}
\centerline{\includegraphics[width=19cm]{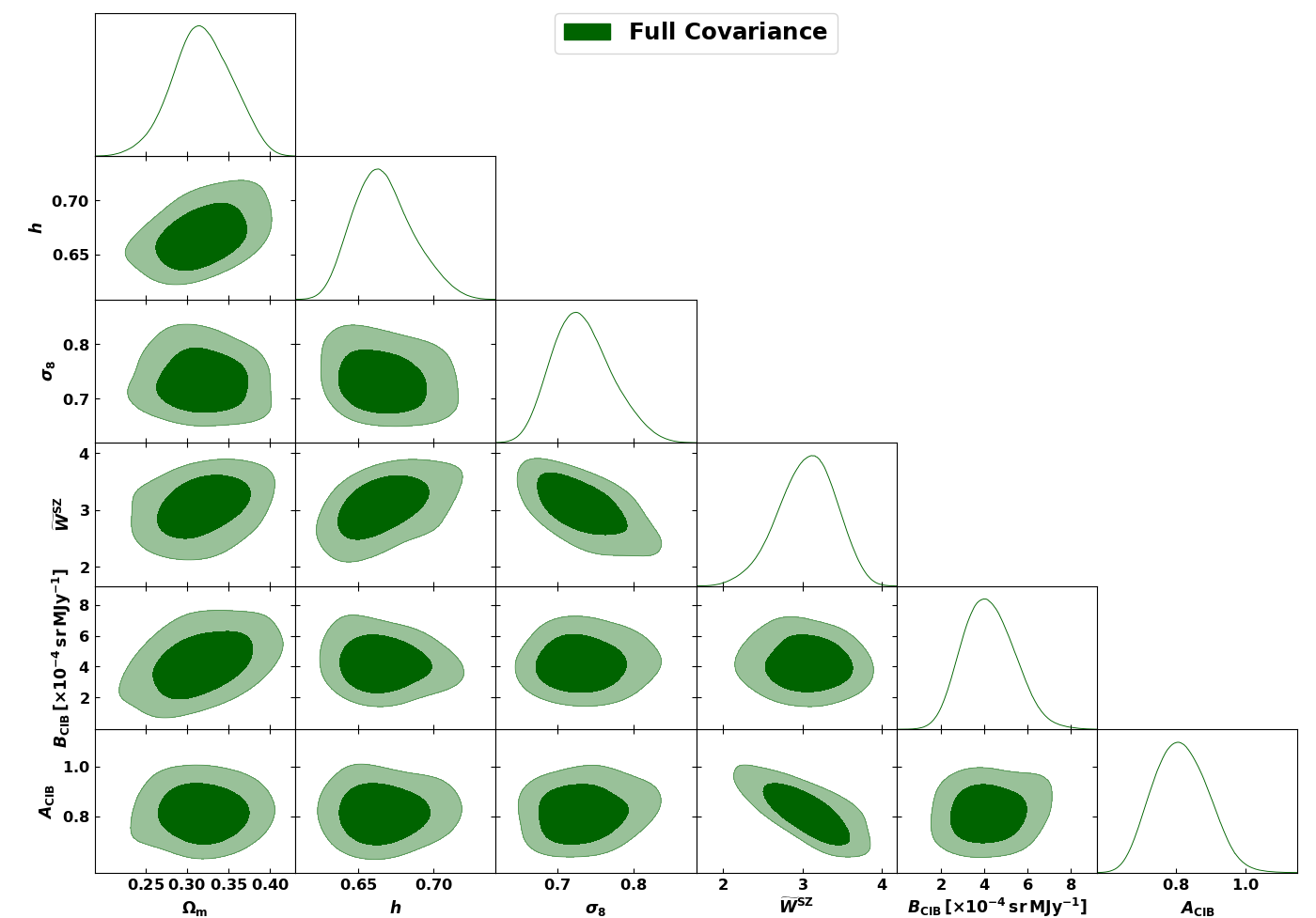}}
	\caption{Posterior distributions from MCMC with the full covariance matrix for the six free parameters in our model. The figure shows the joint constraints for all parameter pairs and the
		marginalized distributions for each parameter along the table diagonal. $\widetilde{W}^{\rm SZ}$ is a dimensionless quantity defined in Equation~(\ref{eq:WSZ_dimensionless}).}
	\label{fig:param_estimatn}
\end{figure*}

Here we compare the theoretical power spectra with the observed spectra. Let the theoretical spectra be $\mathbf{C}^{\rm th}_{\ell}(\Theta)$ and the observed spectra be $\mathbf{C}_{\ell}^{\rm obs}$. The predictions are controlled by the parameter set $\Theta\equiv(\Omega_{\rm m}, h, \sigma_{8}, \widetilde{W}^{\rm SZ}, B_{\rm CIB}, A_{\rm CIB})$, while the observed power spectra $\mathbf{C}_{\ell}^{\rm obs}\equiv (C^{\rm TT}_{\ell}, C^{{\rm T}y}_{\ell}, C^{yy}_{\ell})$. With the full covariance matrix ${\rm Cov}_{\rm tot}$ calculated in Sec.~\ref{ssec:Cov_matrix}, we formulate the $\chi^{2}$ function as
\begin{eqnarray}
\chi^{2}(\Theta) &=& \left(\mathbf{C}_{\ell}^{\rm obs}-\mathbf{C}_{\ell}^{\rm th}\right)\,\left({\rm Cov}_{\rm tot}\right)^{-1} \nonumber \\
& \times & \left(\mathbf{C}_{\ell}^{\rm obs}-\mathbf{C}_{\ell}^{\rm th}\right)^{\rm T}.
\end{eqnarray} 
The likelihood function $\mathcal{L}(\Theta) \sim e^{-\chi^{2}/2}$ and the final posterior probability distribution for the parameters $P(\Theta|d)$ are related by
\begin{eqnarray}
P (\Theta | d) \propto   P(\Theta) \mathcal{L}(d | \Theta ),
\end{eqnarray}
where $P(\Theta)$ is the prior probability function, for which we assume flatness for all parameters within the ranges listed in Table~\ref{tab:estimates}. $\mathcal{L} (d | \Theta)$ is the likelihood function of the data, which includes both contributions from the cosmic variance and noise; it also considers the effect of masks on the data. We then employ the Markov Chain Monte Carlo (MCMC) technique\footnote{We use the \textsc{emcee} package~\citep{emcee} for a defined prior and subsequent chains.} to explore the parameter space. We then scale the constraints on $\sigma_{8}$ by $\Omega_{\rm m}$ to derive $S_8 \equiv \sigma_{8}(\Omega_{\rm m}/0.3)^{0.5}$. The final posterior distributions on our parameter sets are plotted in Figure~\ref{fig:param_estimatn}.

\subsection{Constraints on LSS parameters}
\begin{figure*}
	\centering
\includegraphics[width=15cm]{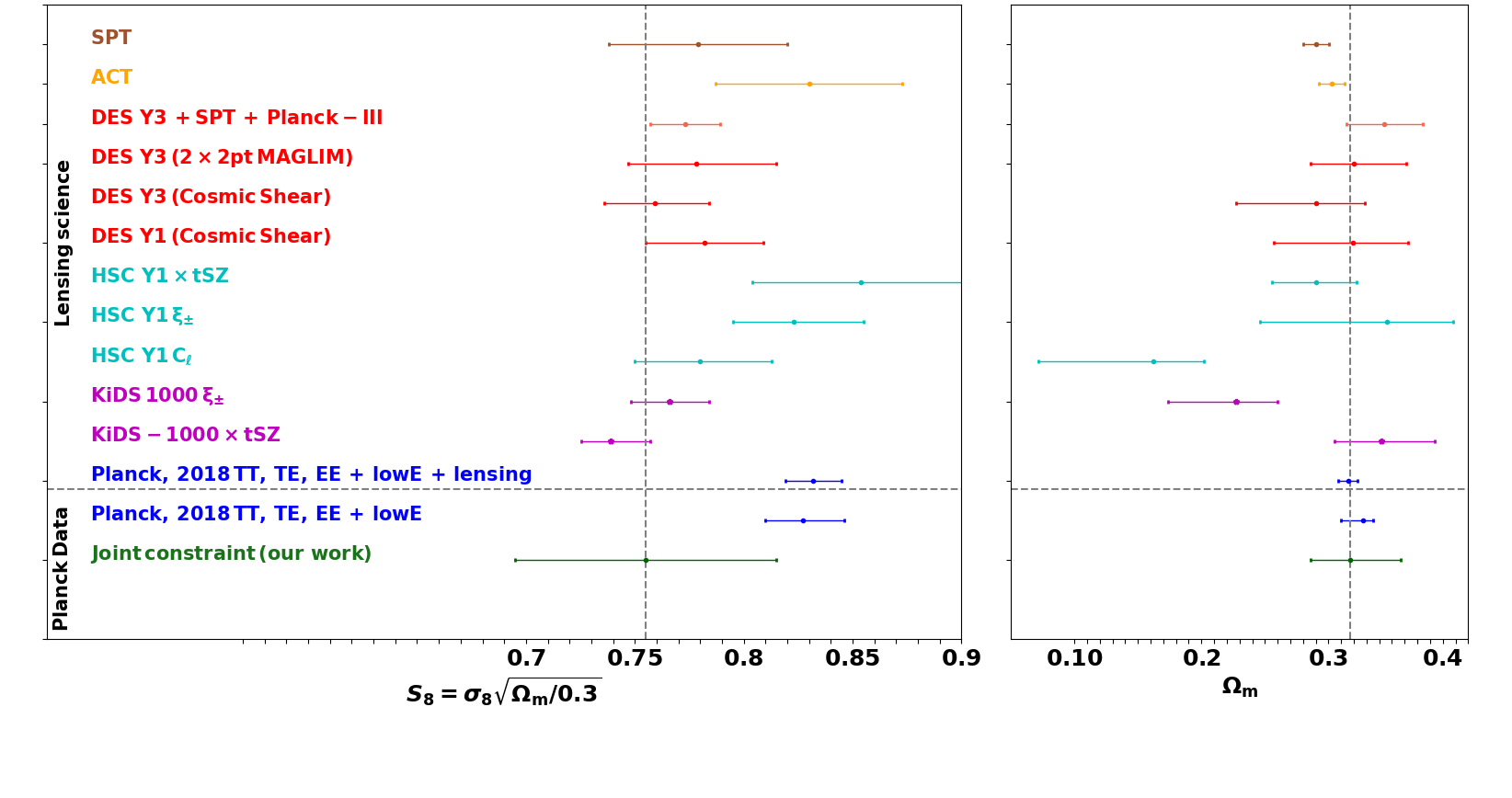}
\caption{{\it Left}: a comparison between the constraints on the amplitude of the mass fluctuations $\sigma_{8}$ scaled by the square root of the matter density $\Omega_{\rm m}$ estimated in this study using angular power spectra and the results from other studies. {\it Right}: a comparison between the constraints on the matter density parameter and the estimates from other studies. These studies include data from SPT-3G~\citep{SPT3G_21}, ACT-DR4~\citep{ACTDR4_20}, DES Y3\,$2\times 2\,{\rm pt}$\,MAGLIM~\citep{DESY3_galGAl}, DES-Y3 cosmic shear~\citep{DESY3_CosmicShear}, DES-Y1 cosmic shear \citep{DESY1_cosmic_shear}, Subaru HSC Y1$\times$tSZ~\citep{Ken-Osato2019}, HSC-Y1\,$\xi_{\pm}$\,cosmic shear~\citep{HSCY1_CosmicShear}, the HSC-Y1 power spectrum~\citep{HSCY1_CLs}, and KiDS-$1000\,\xi_{\pm}$~\citep{KIDS_1000} and KiDS-$1000\times$tSZ \citep{KIDS_1000XtSZ}. We also compare our results with measurements from \citet{Planck2018TT}.}
	\label{fig:S8_comparison}
\end{figure*}

\begin{figure}
	\centering
    \includegraphics[width=3.2in]{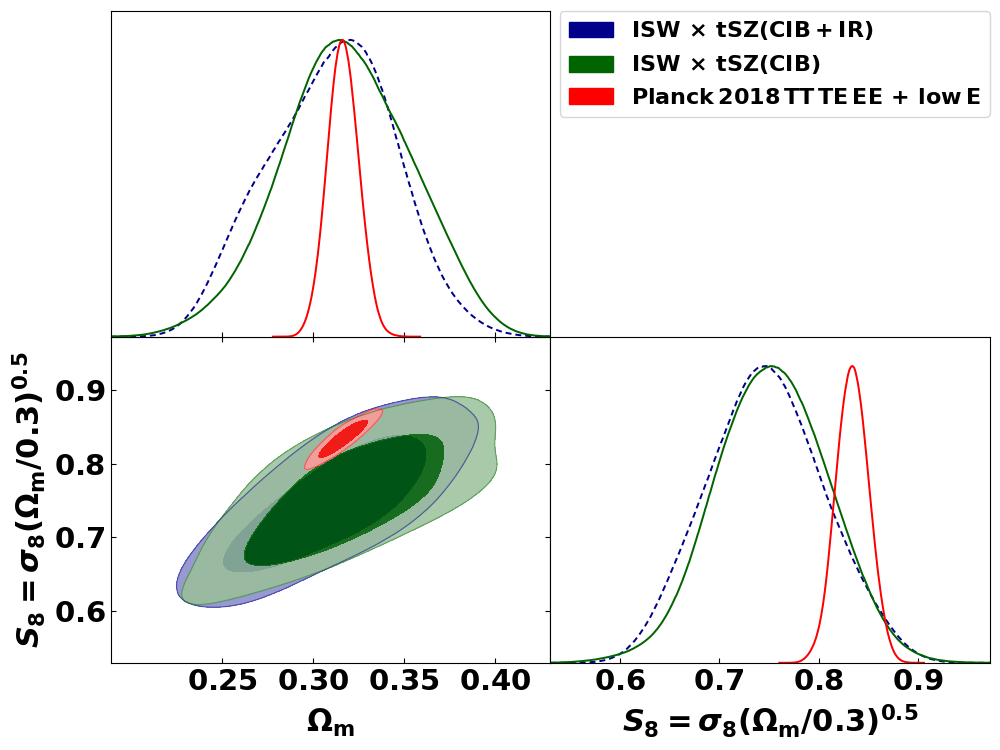}
\caption{Base $\Lambda{\rm CDM}$ model 68\% and 95\% posterior distribution constraint contours on the matter density parameter $\Omega_{\rm m}$ and on the structure amplitude parameter $S_{8}$.}
	\label{fig:planck_like_and_mine}
\end{figure}

Starting with the fundamental parameters, we estimated the matter density parameter, Hubble constant, and the level at which matter clusters (the amplitude of the matter density fluctuations) to be
\begin{eqnarray}
\begin{array}{rcl}
H_{0} & = & 69.7^{+2.0}_{-1.5}\, {\rm km}\,{\rm s}^{-1}\,{\rm Mpc}^{-1}, \\ 
\Omega_{\rm m}& = & 0.317^{+0.040}_{-0.031}, \\
\sigma_{8}& = &0.730^{+0.040}_{-0.037}
\end{array} \  \Bigg \} 68\%\,{\rm \, Joint \,Constraints}\nonumber 
\\
\end{eqnarray}
Additionally, we used the measured correlation functions to derive a constraint on the degenerate combination:
\begin{eqnarray}
S_8 \equiv \sigma_{8}(\Omega_{\rm m}/0.3)^{0.5} = 0.755\pm{0.06} \,\,(68\% \,{\rm CL}).
\end{eqnarray}
Our constraint on $\Omega_{\rm m}$ broadly agrees with previous literature, as shown in Figure~\ref{fig:S8_comparison}. However, we see that the matter density of the Universe, $\Omega_{\rm m}$ is slightly correlated with the tSZ gas parameter.
This is indeed expected, as the amplitude of the tSZ effect (set by the gas parameter) is proportional to the line-of-sight integral of the electron pressure times the electron density along the line of sight. The electron pressure is related to the temperature of the gas, which is determined by the gravitational potential of the cluster and the energy input from various sources, such as accretion shocks and feedback from active galactic nuclei. On the other hand, the electron density is related to the matter density of the gas. Hence, the tSZ effect is a sensitive probe of the underlying matter density of the Universe and should be correlated.
In addition, we see that the tSZ gas parameter is correlated with the Hubble parameter $h$ but anticorrelated with $\sigma_{8}$.
Although the tSZ effect and the Hubble parameter are not directly related to each other, both can be correlated, because they are both sensitive to the properties of galaxy clusters. The tSZ effect in a sample of galaxy clusters can be used to estimate their masses, and distances, which in turn can be used to constrain the Hubble parameter. Hence, $\widetilde{W}^{\rm SZ}$ and $h$ can be correlated. However, there is no direct explanation for the anticorrelation between the tSZ gas parameter and the amplitude of the matter power spectrum on scales of $8\,h^{-1}{\rm Mpc}$. This is because they are sensitive to different aspects of the LSS of the Universe. While $\sigma_8$ characterizes the strength of the clustering of matter on large scales in the Universe, $\widetilde{W}^{\rm SZ}$ relates to the properties of the gas in clusters, so an anticorrelation might be possible. To further prove whether they are both correlated would depend on several factors, including the redshift range and mass range of the specific galaxy clusters contributing to the tSZ effect, as well as the assumed cosmological model, which is beyond the scope of this current work.

We now discuss our constraints on the $S_8$ parameter in comparison with those obtained from other measurements. First, we would like to compare our estimate with \citet{Planck2015_tSZ_paper}, considering the fact that we utilize the same tSZ data. We remind the reader that \citet{Planck2015_tSZ_paper} parameterized the amplitude of matter fluctuation as $S_{8} = \sigma_{8}(\Omega_{\rm m}/0.28)^{3/8}$ to obtain a value of $S_{8} = 0.80^{+0.01}_{-0.03}$ and $S_{8} = 0.90^{+0.01}_{-0.03}$ for a mass bias of 0.2 and 0.4 respectively. Using the same parameterization, our results yield $S_{8} = \sigma_{8}(\Omega_{\rm m}/0.28)^{3/8} = 0.768\pm{0.05}$, a value that is consistent in $0.6\sigma$ (for mass bias\,$=0.2$), and $2.3\sigma$ (for mass bias\,$=0.4$), respectively. In Figure~\ref{fig:S8_comparison}, we further show the comparison of our results with measurements obtained from other analyses, such as the measurements of the E-mode polarization and temperature-E correlation of the CMB from the South Pole Telescope-3G (SPT-3G) 2018 data~\citealt{SPT3G_21}), the Atacama Cosmology Telescope (ACT) Data Release 4 (DR4) data~\citep{ACTDR4_20}, the HSC lensing and cross-correlation with tSZ~\citep{Ken-Osato2019}, and the measurements from the Planck ``TT, TE, EE+lowE'' and ``TT, TE, EE+lowE+lensing'' data~\citep{Planck2018TT}. Our estimate is consistent with the KiDS lensing cross-correlation with tSZ~\citep{KIDS_1000XtSZ}, the Dark Energy Survey (DES) cosmic shear, galaxy clustering, and galaxy\textendash galaxy lensing (Y1 and Y3) results~\citep{DESY3_CosmicShear,DESY3_galGAl}, the HSC lensing cosmic shear~\citep{HSCY1_CosmicShear}, the HSC-Y1 power spectrum~\citep{HSCY1_CLs} and SPT-3G~\citep{SPT3G_21}, within the estimated errors. However, we prefer a slightly lower value compared to the CMB results. 

In contrast to the usual banana-shaped contour, which characterizes the constraints between the $S_8$ and $\Omega_{\rm m}$ degeneracy from the cosmic shear and shear\textendash tSZ cross correlation~\citep{DES_BAO_BBN}, the shape of our constraints is consistent with the Planck TT, TE, and EE+ lowE measurements~\citep{Planck2018TT} within $1.1\sigma$, as shown in Figure~\ref{fig:planck_like_and_mine}. Indeed, our analysis combined the tSZ with the ISW tracer which probes the late-time Universe. Therefore, such a small difference could arise from the contribution introduced by the late-Universe probe in our work.

\subsection{Constraints on the gas parameter}   
\label{sec:constrain -WSZ}
The tSZ gas parameter is estimated as the joint constraint on the product of the mean electron density $\bar{n}_{\rm e}$, electron temperature $T_{\rm e}$, and gas bias $b_{\rm gas}$ at redshift $z=0$. We parameterize the product as $\widetilde{W}^{\rm SZ}$ (via Equation~(\ref{eq:WSZ_dimensionless})) and obtained a value of\footnote{For $C^{yy}_{\ell}$-only constraint on $\widetilde{W}^{\rm SZ}$ parameter, please refer to Appendix~\ref{App:gas} and Figure~\ref{fig:Compare_WSZ}.} 
\begin{equation}
   \widetilde{W}^{\rm SZ} \equiv b_{\rm gas} \left( \frac{T_{\rm e}}{0.1\,{\rm keV}}  \right ) \left( \frac{\bar{n}_{\rm e}}{1\,{\rm m}^{-3}} \right) = 3.09^{+0.320}_{-0.380}. \label{eq:WSZ_constraint}
\end{equation}
This result is consistent with the value estimated by~\citet{V.Waerbeke2014} within $1.7 \sigma$. Let us remark that, as explained above, in our analysis we fixed the $n_{\rm s}$, $\tau$, and $A_{\rm s}$ parameters. This choice was taken because our parameter space could be too complicated for our likelihood. However, it is worth noting that, whereas the $n_{\rm s}$ and $A_{\rm s}$ uncertainties are very small compared to the ones derived for our parameters, $\tau$ is constrained at around $12\%$ by \citet{Planck2018TT}, similar to the one obtained for $\widetilde{W}^{\rm SZ}$. Taking into account that $\widetilde{W}^{\rm SZ}$ is proportional to $n_{\rm e}$ and that this is also proportional to $\tau$, we can derive a rough estimation of how much our uncertainty on $\widetilde{W}^{\rm SZ}$ could grow by if $\tau$ would be marginalized instead of fixed. In that case, our error will grow from the actual $12\%$ to $18\%$.

\begin{figure}
	\centering
    \includegraphics[width=3.3in]{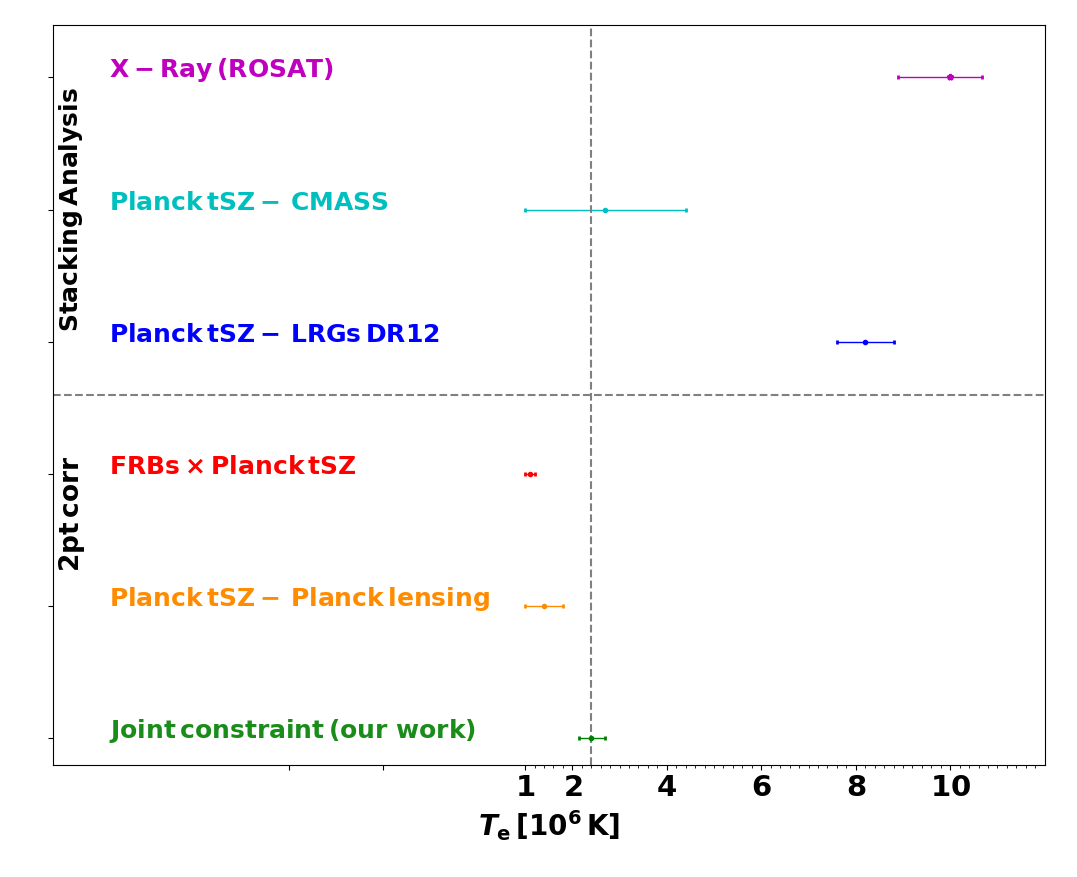}
\caption{Comparison of the estimate on baryon temperature from two-point correlation studies \citep[FRBs $\times$ Planck\,tSZ;][]{FRB_tSZ}, stacking analyses (Planck tSZ-LRGs DR12\textendash~\citealt{Tanimura2019}; Planck tSZ-CMASS\textendash~\citealt{de-Graaff2019}; Planck tSZ-Planck lensing\textendash~\citealt{Tanimura20}), and X-ray observations \citep[X-Ray\, ROSAT;][]{Wang1993}. Notice that the temperature values derived from different measurements correspond to cosmic structures on different scales; hence, this plot is only intended to provide a comparison between their orders of magnitude, which are found to be in broad agreement.}
	\label{fig:Compare_Te}
\end{figure}

We can further derive an estimate on the electron temperature for today by adopting the numerical values of $b_{\rm gas}$ and $\bar{n}_{\rm e}$ from~\citet{Refregier2000} and~\citet{Seljak2001}, as $4 \leq b_{\rm gas} \leq 9$. Taking $b_{\rm gas}=6$ and $\bar{n}_{\rm e}=0.25\,{\rm m}^{-3}$, we found 
\begin{eqnarray}
T_{\rm e}= (2.40^{+0.250}_{-0.300}) \times 10^{6}\,{\rm K},
\end{eqnarray}
which is consistent with the values obtained from~\citet{de-Graaff2019}, in $0.18\sigma$, with the expected WHIM temperature. Note that that analysis in the work of \citet{de-Graaff2019} was entirely interpreted using halo modeling and hence was able to resolve individual halos, thereby providing information on filaments in Galaxy clusters up to small scales. Now we provide a comparison of our estimate on the electron temperature with other studies in Figure~\ref{fig:Compare_Te}.
\begin{figure}
	\centering
    \includegraphics[width=3.3in]{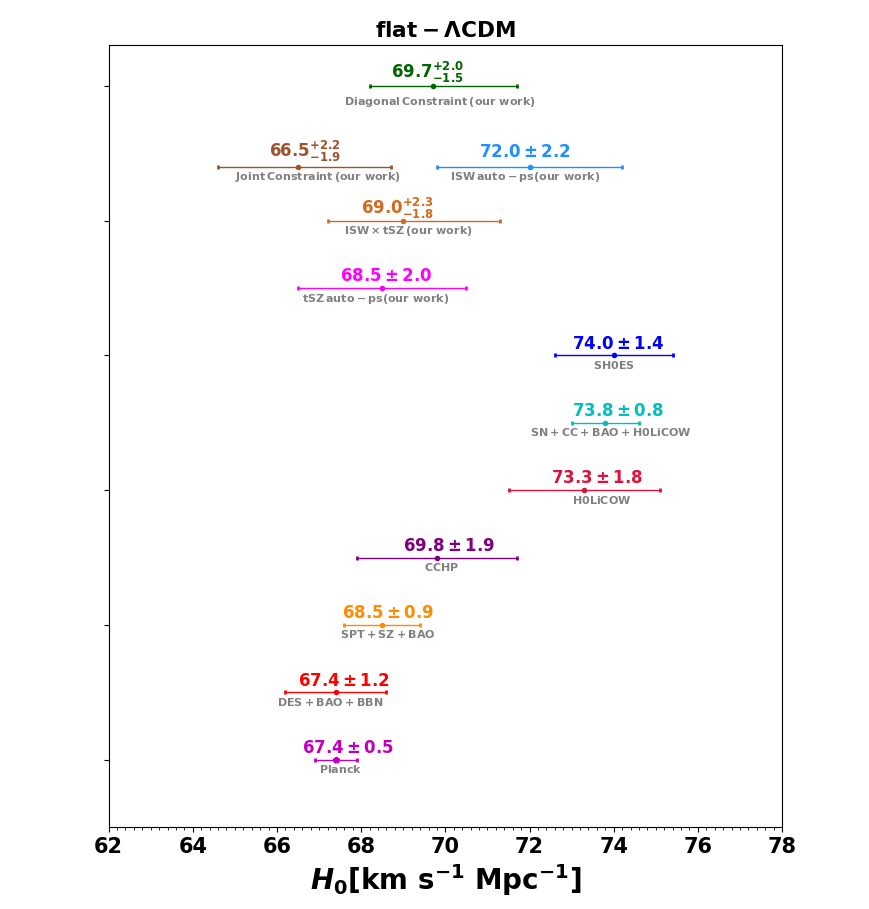}
\caption{A comparison of the early- and late-time measurements of the Hubble constant. These studies include Planck data~\citep{Planck2018TT}, DES Year 1, BAO, and BBN data \citep[DES+BAO+BBN;][]{DES_BAO_BBN}, SPT-SZ and BAO \citep[SPT-SZ+BAO;][]{Addison18}, CCHP \citep{CCHP19}, lensed quasars data \citep[H0LiCOW;][]{H0LiCOW20}, the SN, CC, BAO, and H0LiCOW lenses sample \citep[SN+CC+BAO+H0LiCOW;][]{Bonilla21}, and Hubble Space Telescope observations \citep[SH0ES;][]{Riess19}.}
\label{fig:H0_comparison}
\end{figure}

We found our result in broad agreement with previous studies from numerical simulations and observational analyses. Hydrodynamic simulations~\citep{Ostriker1999,Ostriker2001} showed that the diffuse baryons locked up in the intergalactic medium (IGM) are heated up to $T\sim 10^{5}$\textendash $10^{7} {\rm K}$, which suggests that a substantial fraction of the ``missing baryons'' in the Universe are in the phase of the warm\textendash hot IGM that traces dark matter.

We now compare our results with other observational analyses. One should notice that different studies probe the LSS tracers with different densities and environments; hence, it is not very meaningful to directly compare the estimates of gas temperature from different measurements. Nonetheless, the cross-correlation analysis usually receives contributions from the two-halo term, which corresponds to the correlated gas on large scales. In contrast, stacking analyses are usually sensitive to the particular small-scale configurations of individual systems, which usually correspond to high-density regions. Therefore, in Figure~\ref{fig:Compare_Te}, we separate the two-point correlation measurements of the gas temperature from stacking analyses. 

In the same category of two-point correlation constraints, \citet{FRB_tSZ} cross-correlated FRBs with the tSZ map and estimated an IGM temperature of $T_{\rm e} = (1.1 \pm {0.1}) \times 10^{6}\,{\rm K}$. The cross-correlation between the tSZ maps from the Planck data and the lensing convergence map from the Canada\textendash France\textendash Hawaii Lensing Survey~\citep{CHFTLenS} showed that the diffuse baryons, which are responsible for the lensing\textendash SZ cross-correlation, have temperatures $10^{5}$\textendash $10^{7}\,{\rm K}$~\citep{V.Waerbeke2014,Ma2015}. \citet{Santos15} performed a follow-up analysis using the foreground-cleaned CMB maps in cross-correlation with the Two-Micron All Sky Redshift Survey (2MASS) of galaxies and obtained a fraction of the baryons at temperatures in the range $10^{4.5}$-$10^{7.5}\,{\rm K}$, consistent with \citet{V.Waerbeke2014} at $z\sim0.5$. The temperature values found from these studies are in broad agreement with our current study.

As for other analyses, \citet{Eckert2015} investigated the A2744 cluster over a physical length of $8\,{\rm Mpc}$ using X-ray observations and found out that the filamentary structures of gas are in a plasma temperature range of $T = 1 \times 10^{7}\,K$\textendash $2\times 10^{7}$K for the various filaments, after discarding and eliminating coincidental X-ray point sources. Similarly, \citet{Bulbul16} did a thorough analysis of the A1750 cluster using Suzaku and Chandra X-ray observations and Multiple Mirror Telescope optical observations, to report that in both the filaments and off-filament directions, the gas mass is in agreement with the cosmic baryon fraction at $R_{200}$. In addition, for stacking analyses, \citet{Tanimura2019} stacked the luminous red galaxy samples from the DR12 catalog of SDSS on Planck tSZ map to search for warm gas filamentary structure. They obtained a constraint on the mean electron-density-weighted temperature as $\sim (8.2 \pm 0.6) \times 10^6\,{\rm K}$. With X-ray observations, \citet{Tanimura_Xray22} obtained an average temperature of $(1.2 ^{+0.35}_{-0.23}) \times 10^7\,{\rm K}$. \citet{Alvarez18} also obtained a measurement of the global projected filament temperature to an estimate of $T_{\rm e} = (5.2^{+1.03}_{-0.64}) \times 10^{7}\,{\rm K}$, using measurements from the Chandra and XMM-Newton X-ray observations of the A3391/A3395 intercluster filament, an estimate that is higher than other measurements, including our results, but compatible with \citet{Tanimura_Xray22} in the $3.8\sigma$ CL. 

These single-system stacking analyses used specific tracers to infer the WHIM properties. We stress that our result is somehow in between the stacking and cross-correlation results, as shown in Figure~\ref{fig:Compare_Te}. We think this fact deserves further studies with numerical simulation and the kinematic Sunyaev\textendash Zeldovich (kSZ) effect ~\citep{kSZ,chm15,Ma2017,Schaan2021}. In terms of breaking degeneracy in Equation~(\ref{eq:WSZ_constraint}), the kSZ effect is solely dependent on the electron density and peculiar velocity of the cluster, so if the latter can be inferred by another method (e.g. reconstructing the velocity field from the redshift-space density field;~\citet{chm15,Planck2016_XXXVII}), the electron density can be measured. In this way, the degeneracy between the electron density and temperature can be broken and one can provide an independent measurement of the gas temperature (see, e.g.~\citealt{Schaan2021}). Other relevant measurements that have probed WHIM properties include but are not limited to~\citet{Wang1993}, \citet{Werner2008}, \citet{Nevalainen19} and~\citet{Erciyes23}.

We now consider the constraints on the foreground parameters. We quantify the level to which the clustered CIB contributes to the observed auto-power spectrum for $yy$ and find a value of
\begin{eqnarray}
A_{\rm CIB} = 0.81^{+0.086}_{-0.079}.     
\end{eqnarray}
We remark that the anticorrelation between $A_{\rm CIB}$ and $\widetilde{W}^{\rm SZ}$ seen in Figure~\ref{fig:param_estimatn} is expected, because the tSZ effect and the CIB arise from different physical mechanisms. The CIB is thought to be produced by star formation in galaxies, which is more likely to occur in regions of high gas density, while the tSZ effect occurs in regions of low gas density. Also, the redshift and mass dependences of the tSZ effect and the CIB are different. While the tSZ comes from more massive clusters, CIB is stronger for less massive objects. Therefore, an anticorrelation is possible between the two parameters.

Similarly, we stress that the clustered CIB could also contribute to the cross-correlation function, as reported in ~\citet{Makiya2018}. Therefore, we parameterized such a contribution in this analysis with $B_{\rm CIB}$. This parameter gauges the contamination from the clustered CIB to the ${{\rm T}y}$ cross-correlation. We obtained the best-fitting value in the order of $\sim 10^{-4}\,\text{sr}\,\text{MJy}^{-1}$ and show in Figure~\ref{fig:spectra} its contribution to the correlation of the Planck ISW with the Compton parameter map. This contamination is dominant on low -$\ell$, high $z$ (${\ell_{\rm eff}} \leq 30$). 

\subsection{Hubble constant measurement}
\label{sec:Hubble_tension}
The Hubble constant is a fundamental cosmological parameter that quantifies the expansion rate of the Universe. In recent years, the tension on the value of $H_0$ has risen between early (global) and late (local) Universe measurements (see Figure~\ref{fig:H0_comparison} for the comparison). The early Universe measurement from the CMB radiation prefers a somewhat lower value, as $H_0 = 67.4 \pm 0.5\, {\rm km}\,{\rm s}^{-1}\,{\rm Mpc}^{-1}$~\citep{Planck2018TT}. The DES clustering, baryon acoustic oscillation (BAO), and Big Bang Nucleosynthesis (BBN) measurements also give a lower value $H_0 = 67.4 \pm 1.2\, {\rm km}\,{\rm s}^{-1}\,{\rm Mpc}^{-1}$~\citep{DES_BAO_BBN}. 
In contrast, late-time observables give a higher value than these. The standard distance ladder from Cepheid variables gives $H_0 = 74.0 \pm 1.4\, {\rm km}\,{\rm s}^{-1}\,{\rm Mpc}^{-1}$~\citep{Riess19}, and the strong-lensing time-delay distances (H0LiCOW team) give $H_0 = 73.3 \pm 1.8\, {\rm km}\,{\rm s}^{-1}\,{\rm Mpc}^{-1}$~\citep{H0LiCOW20}. In addition, a combination of supernova type Ia (SN), cosmic chronometers (CC), and BAO data with the time-delay cosmography from H0LiCOW were used to obtain a value of $H_0 = 73.8 \pm 0.8\, {\rm km}\,{\rm s}^{-1}\,{\rm Mpc}^{-1}$~\citep{Bonilla21}. However, the $H_0$ value obtained by the Carnegie\textendash Chicago Hubble Program (CCHP), based on the Tip of the Red Giant Branch (TRGB) measurements in the Large Magellanic Clouds, falls between the early- and late-time measurement, yielding $H_0 = 68.9\pm 1.9\, {\rm km}\,{\rm s}^{-1}\,{\rm Mpc}^{-1}$~\citep{CCHP19}.

In this work, to study the robustness of our Hubble constant measurement, we carefully examine the $H_{0}$ value using individual likelihoods and joint probes. The individual likelihood just uses the specific covariance blocks in Figure~\ref{fig:correlation}, calculated via Equation~(\ref{eq:Cell_tot}). By using tSZ data only (the middle block in Figure~\ref{fig:correlation} for its covariance) and ISW data only (the lower left block), we obtain
\begin{eqnarray}
H_0 &=& 68.5 \pm{2.0} \, {\rm km}\,{\rm s}^{-1}\,{\rm Mpc}^{-1}\,\, ({\rm tSZ}), \nonumber \\ 
H_0 &=& 72.0 \pm{2.2} \, {\rm km}\,{\rm s}^{-1}\,{\rm Mpc}^{-1} \,\,({\rm ISW}),
\label{eq:H0_individual}
\end{eqnarray}
where for tSZ we marginalized over the ($\Omega_{\rm m}$, $\sigma_{8}$, $\tilde{W}^{\rm SZ}$, $B_{\rm CIB}$, $A_{\rm CIB}$) parameters, and for ISW we marginalized only the ($\Omega_{\rm m}$, $\sigma_{8}$) parameters. One can see that the tSZ-only data make the $H_{0}$ value consistent with the Planck result within the error range, whereas the ISW data make the $H_{0}$ constraint closer to the ``SH0ES'' local measurements~\citep{Riess19}, ``SN+CC+BAO+H0LiCOW'' from~\citet{Bonilla21}, and H0LiCOW measurement (\citealt{H0LiCOW20}; see Figure~\ref{fig:H0_comparison}).
By using the tSZ\textendash ISW cross-correlation likelihood (using the ${\rm Cov}^{{\rm T}y, {\rm T}y}$ covariance; i.e., the upper right block of Figure~\ref{fig:correlation}), we obtained
\begin{eqnarray}
H_0 = 69.0 ^{+2.3}_{-1.8}\, {\rm km}\,{\rm s}^{-1}\,{\rm Mpc}^{-1}\,\, ({\rm tSZ-ISW}), \label{eq:H0_cross}
\end{eqnarray}
which error bar overlaps with both single-likelihood cases and is consistent with the tSZ-only case at the $\lesssim 0.5\sigma$ CL.

We now combine the datasets and constrain $H_0$. We first used the full cross-correlation data by adding cross-covariance terms in Equation~(\ref{eq:Cell_tot}), i.e. using all the blocks in Figure~\ref{fig:correlation}, and obtained
\begin{eqnarray}
H_0 = 66.5^{+2.0}_{-1.9}\, {\rm km}\,{\rm s}^{-1}\,{\rm Mpc}^{-1} \,\,({\rm Joint}).
\label{eq:H0_allcombined}
\end{eqnarray} 
This result is in excellent agreement with the values obtained by Planck~\citep{Planck2018TT} and that obtained by the DES clustering, BAO, and BBN measurement~\citep{DES_BAO_BBN} in the $\lesssim 0.5 \sigma$ CL. However, this joint constraint sits outside the range of individual constraints, and deviates from the ISW-only constraint by more than $1\sigma$. As a sanity check, we added the offdiagonal cross-covariance matrix one after the other to the diagonal covariance and found that the driving force of this low-$H_0$ value is the correlated cosmic variance from the TT to T$y$ statistics. In other words, all the different combinations except the cross-covariance ${\rm Cov}^{{\rm TT}, {\rm T}y}$ provide consistent results with the two individual constraints. Therefore, instead of combining all covariance blocks, if we only use the diagonal blocks of the covariance matrices (the TT, $yy$, and T$y$ autocorrelations), we obtain
\begin{eqnarray}
H_0 = 69.7 ^{+2.0}_{-1.5}\, {\rm km}\,{\rm s}^{-1}\,{\rm Mpc}^{-1}  \,\,({\rm Diagonal}), \label{eq:H0_diagonal}
\end{eqnarray} 
which sits in between the ISW-only and tSZ-only constraints and is close to the tSZ and ISW cross-correlation-only constraint (the green data point in Figure~\ref{fig:H0_comparison}). This result is also in excellent agreement with the CCHP measurement ($<0.1 \sigma$;~\citealt{CCHP19}), the Planck result ($\sim 1\sigma$; \citealt{Planck2018TT}) and that obtained by the DES clustering, BAO, and BBN measurement ($\sim 1.2 \sigma$;~\citealt{DES_BAO_BBN}), respectively. It is also compatible with the SH0ES~\citep{Riess19} and H0LICOW~\citep{H0LiCOW20} measurements in $\sim 1.8 \sigma$ and $\sim 1.4 \sigma$, respectively. Comparing Equation~(\ref{eq:H0_allcombined}) with Equations~(\ref{eq:H0_diagonal}), (\ref{eq:H0_individual}), and (\ref{eq:H0_cross}) (also in Figure~\ref{fig:H0_comparison}), we conclude that the low value of the joint probes may be due to some unaccounted systematics in the TT\textendash T$y$ covariance block. Therefore, instead of quoting the ``all-combined'' constraint on $H_0$, we quote the joint constraint of the $H_{0}$ value obtained by using the diagonal blocks of the covariance matrix, i.e. Equation~(\ref{eq:H0_diagonal}). For completeness of different scenarios, we also show the  $H_{0}$ value obtained using the full covariance case and the individual constraints in Figure~\ref{fig:H0_comparison}.

We also notice that recently \citet{Salvatore23} reported that the Hubble tension can be removed in the framework of $\Lambda$CDM in terms of the cosmological lookback time (the redshift at which the measurements are performed). No definitive conclusion has been reached about the value of $H_0$, which provides a strong reason to suspect the existence of physics beyond the $\Lambda$CDM model~\citep{Dai2020,Perivolaropoulos22}. Future CMB experiment such as the Simon Observatory~\citep{Ade2019} and CMB-S4~\citep{CMB-S4}, combined with the new generation of galaxy surveys, such as Euclid and LSST, are expected to reach a precision of $\sim 0.15\%$ in the $H_0$ estimate~\citep{Valentino21}, thus providing a stringent test for the adopted cosmological models.

\section{Conclusions}
\label{sec:conclusions}
In this paper, we have provided a novel constraint on the gas parameter on large scales using the cross-correlation of the ISW with the tSZ effect. There have been several theoretical studies on the benefits of cross-correlating the ISW with the tSZ effect~\citep{ISW_tSZ_02,ISW_tSZ_011,Creque-Sarbinowski16}, but the actual measurement with the Planck data has not been done before. In this work, we measured the tSZ\textendash ISW cross-correlation power spectrum by using the Planck ISW data and the tSZ Compton-$y$ map from the Planck NILC algorithm. We then use the cross-correlation data to constrain fundamental cosmological parameters $(\Omega_{\rm m}, h, \sigma_{8})$ and the gas properties on very large scales, via the parameter $\widetilde{W}^{\rm SZ}= b_{\rm gas} \left(T_{\rm e}/0.1\,{\rm keV}  \right ) \left(\bar{n}_{\rm e}/1\,{\rm m}^{-3} \right)$. Our estimates on $\Omega_{\rm m}$ and $\sigma_{8}$ enabled us to estimate the amplitude of matter density fluctuations $S_{8}\equiv \sigma_{8}(\Omega_{\rm m}/0.5)^{0.5}$. We also calculated the theoretical
power spectrum based on large-scale gas bias models to account for the measured spectra~\citep{V.Waerbeke2014}. We used MCMC methods to fit the model parameters and derived large-scale constraints on the cosmological and gas parameters. 

For the measurement part, we employed the MASTER (pseudo-C$_{\ell}$) approach encoded in the \texttt{NaMaster} code to measure the auto- and cross-correlation angular power spectra. Our measured ISW and tSZ autocorrelation power spectra ($C^{\rm TT}_{\ell}$ and $C^{yy}_{\ell}$) are consistent with previous measurements (ISW~\textendash see Figure~7 of~\citealt{Rahman16}; tSZ\textendash see the results in~\citealt{Makiya2018, Bolliet18,Ken-Osato2019,Tanimura22,Ibitoye22}). The important result is the first detection of the ISW\textendash tSZ cross-correlation power spectrum. Here we simulated $900$ ISW maps to quantify the covariance of correlation with tSZ data. With the $100$ simulated ISW maps being cross-correlated with true tSZ data, we quantify that the true ISW-tSZ cross-correlation is detected at the $3.6\sigma$ CL., with its amplitude and shape compatible with the theoretical prediction in~\citet{Creque-Sarbinowski16}~(see, e.g., their Figure~1). To maximize the statistical information in our measurement we combined the power spectra into a vector and then fit for the cosmological and astrophysical parameters. We explored the parameter space with an MCMC method, using the \texttt{Python} \textsc{emcee} package \citep{emcee} to determine the maximum likelihood of our data, and obtained posterior probability distributions on all parameters. For a sanity check, we also used another independent package, \textsc{ultranest} \citep{ultranest,ultranest2}, to explore the parameter space and found its results compatible with the ones obtained using \textsc{emcee} package (see Figure~\ref{fig:compare_constraints} for such a comparison). We then adopted the \textsc{GetDist} \texttt{Python} software detailed in~\citet{Antony19} to plot the final distributions in Figure~\ref{fig:param_estimatn}.

We show the constraints obtained on $\Omega_{\rm m}$, and $\sigma_{8}$ in Figure~\ref{fig:S8_comparison}.  Additionally, we compared different $H_0$ measurements from the literature and summarized them in Sec.~\ref{sec:Hubble_tension} and Figure~\ref{fig:H0_comparison}. Our cross-correlation-data estimated $H_0$ value is in between the CMB measured value and the local distance ladder measurement value (SH0ES) and is as such more consistent with the result from the TRGB measurement~\citep{CCHP19}. On the other hand, our diagonal constraints yield a value of $H_0 =69.7^{+2.0}_{-1.5}\, {\rm km}\,{\rm s}^{-1}\,{\rm Mpc}^{-1}$, a value that sits between the global and local measurements of the Hubble constant, while the joint analysis prefers a value of $H_0 =66.5^{+2.2}_{-1.9}\, {\rm km}\,{\rm s}^{-1}\,{\rm Mpc}^{-1}$, a value that is consistent with Planck.

We obtained a joint fit of the gas bias, electron temperature, and electron number density as $\widetilde{W}^{\rm SZ}= b_{\rm gas} \left(T_{\rm e}/0.1\,{\rm keV}  \right ) \left(\bar{n}_{\rm e}/1\,{\rm m}^{-3} \right)=3.09^{+0.320}_{-0.380}$. We used the model amplitude, and an assumed value of $b_{\rm gas}$ and $\bar{n}_{\rm e}$, to estimate the electron temperature as $T_{\rm e}=(2.40^{+0.250}_{-0.300}) \times 10^{6}\,{\rm K}$. Our cross-correlation signal would then account for all the missing baryons that lie within the temperature range $10^{5}$\textendash$10^{7}\,{\rm  K}$ if we assume that they were all located in the WHIM component.

Last, we also estimated the CIB contamination to the cross-correlation function as $B_{\rm CIB}$, while $A_{\rm CIB}$ dictates the foreground amplitude of the CIB contamination in the Compton parameter map. The results are $B_{\rm CIB}\times 10^{4}=4.0^{+1.200}_{-1.100}$ and $A_{\rm CIB}=0.81^{+0.086}_{-0.079}$.
We also derived a constraint on the amplitude of the matter fluctuations on an $8h^{-1}\,{\rm Mpc}$ scale, $S_8=\sigma_{8}(\Omega_{\rm m}/0.3)^{1/2}=0.755\pm{0.060}$. Our result is in broad agreement with the values obtained in the KiDS measurement and our error estimate with that obtained in the DES cosmic shear~Y1 and Y3 results. In general, we show in Figure~\ref{fig:S8_comparison} a comparison of our result with different data and methods (\citealt{DESY1_cosmic_shear,Ken-Osato2019,HSCY1_CLs,HSCY1_CosmicShear,KIDS_1000,DESY3_galGAl,DESY3_CosmicShear,KIDS_1000XtSZ}).

Upcoming surveys like LSST~\citep{LSST_DESC1}, Euclid~\citep{Euclid-collaboration}, and CMB Stage-4 experiments~\citep{CMB-S4}, with increased redshift depth and sky coverage, will provide a better measurement of the ISW than current observational data. The better angular resolution of CMB-S4 and the higher-precision measurement of the ISW will enable the extension of this correlation up to high-$\ell$ modes, which have the promise to reach a much higher precision of the cross-correlation measurement (\citealt{ISW_tSZ_02} forecasted a $\sim 60 \sigma$ detection). In parallel, high-resolution measurements would require a full halo model to account for the diffuse baryon components in halos with different masses. Combining sophisticated modeling with accurate data will constrain astrophysical and cosmological parameters to higher precision. 


\section*{Acknowledgements} 
We thank Simeon Bird, Oluwayimika Ibitoye, Guo-Jian Wang, Yichao Li, and Ziang Yan for the helpful discussion. A.I., Y.Z.M., A.A., and A.B. are funded by the research program ``New Insights into Astrophysics and Cosmology with Theoretical Models Confronting Observational Data'' of the National Institute for Theoretical and Computational Sciences of South Africa. A.I. acknowledges the support of the National SKA Program of China with grant No.~2022SKA0110100, and the Alliance of International Science Organizations, grant No. ANSO-VF-2022-01. Y.Z.M. acknowledges the support of the National Research Foundation with grant No.~150580. P.V. thanks the Spanish Agencia Estatal de Investigación (AEI, MICIU) for the financial support provided under the projects with references PID2019-110610RB-C21, ESP2017-83921-C2-1-R, and AYA2017-90675-REDC, co-funded with EU FEDER funds, and acknowledges support from Universidad de Cantabria and Consejería de Universidades, Igualdad, Cultura y Deporte del Gobierno de Cantabria, via the "Instrumentación y ciencia de datos para sondear la naturaleza del universo" project, as well as from Unidad de Excelencia María de Maeztu (MDM-2017-0765). D.T. acknowledges financial support from the XJTLU Research Development Fund (RDF) grant with the number RDF-22-02-068. W.M.D. acknowledges the support from the ``Big Data for Science and Society'' UKZN Research Flagship.
X.L. acknowledges the support of the Ministry of Science, and Technology (MoST) inter-government cooperation program China\textendash South Africa Cooperation Flagship Project 2018YFE0120800. We thank R. Bel{\'e}n Barreiro for providing us with the Planck ISW simulations.

\section*{Software and data}

For the analysis presented in this manuscript, we made use of the following software:  {\sc HEALPix}~\citep{Gorski2005}, {\sc MASTER}~\citep{Hivon02}, \texttt{NaMaster}~\citep{NaMaster}, {\sc emcee}~\citep{emcee}, {\sc ultranest}~\citep{ultranest,ultranest2}, and {\sc GetDist}~\citep{Antony19}.
The data underlying this article are publicly available. The Planck Compton 
parameter map, the Integrated Sachs\textendash  Wolfe map, the Galactic, and the point-source masks are available at the Planck Legacy Archive at \url{http://pla.esac.esa.int/pla}. The cross-correlation products and their covariance matrices can be shared on request to the corresponding author.

\appendix

\section{Gas Parameter}
\label{App:gas}
\begin{figure}
	\centering
    \includegraphics[width=3.3in]{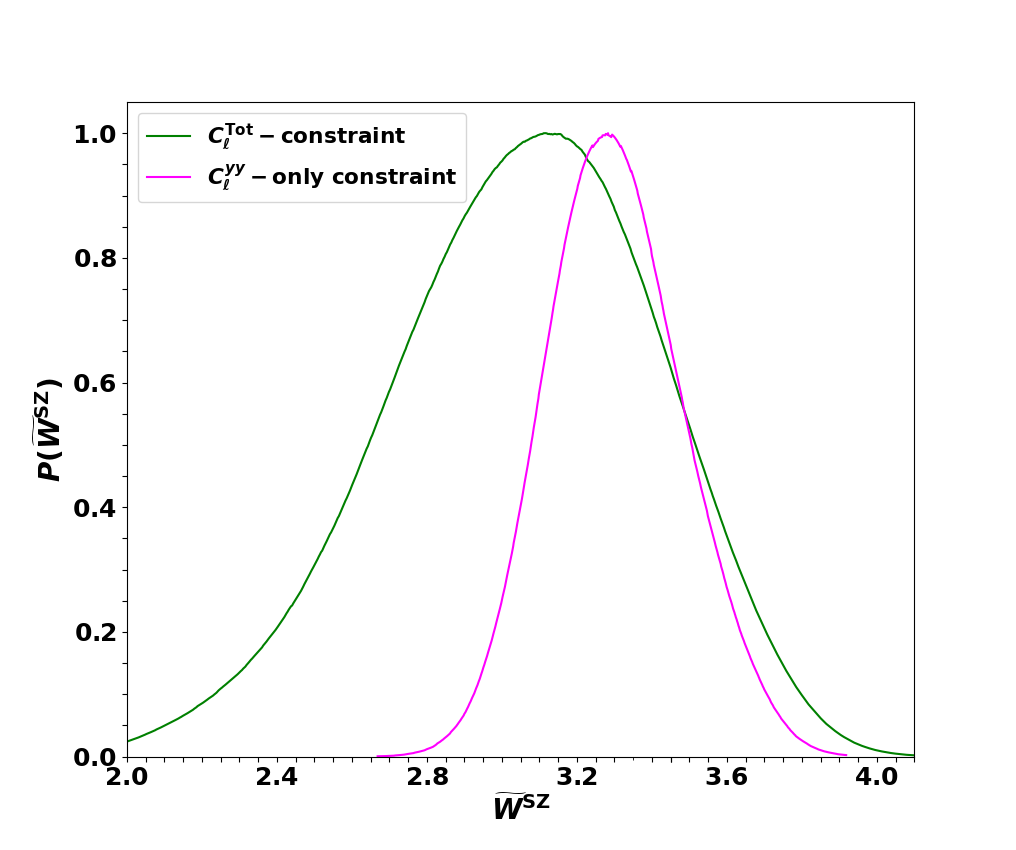}
\caption{Comparison of the estimates on the gas parameter $\widetilde{W}^{\rm SZ}$ (Equation~(\ref{eq:WSZ_dimensionless})) using the $yy$-auto power spectrum and the joint analysis.}
	\label{fig:Compare_WSZ}
\end{figure}

The gas parameter, $\widetilde{W}^{\rm SZ}$, which is defined as a product of the mean electron density $\bar{n}_{\rm e}$, electron temperature $T_{\rm e}$, and gas bias $b_{\rm gas}$, at redshift $z=0$, is a quantity that directly sets the amplitude of the $yy$-auto power spectrum. Figure~\ref{fig:Compare_WSZ} shows the marginalized likelihoods of $\widetilde{W}^{\rm SZ}$ by using only the $C^{yy}_{\ell}$-only and the joint spectra ($C^{\rm Tot}_{\ell}$, i.e. Equation~(\ref{eq:Cell_tot})). One can see that $\widetilde{W}^{\rm SZ} = 3.29^{+0.28}_{-0.24}$ for $C^{yy}_{\ell}$-only and $\widetilde{W}^{\rm SZ} = 3.09^{+0.32}_{-0.38}$ for the joint spectra. These two estimates are consistent within $0.5\sigma$ and the difference in the constrained value may not be due to the fact that a higher value of the gas parameter is needed to reproduce the $yy$-auto power spectrum, but that the tSZ is more sensitive to the abundance of cosmic gas (baryons) on large scales.

\label{sec:WSZ}

\section{Consistency Check on Parameters}
\label{App:consistency}
\begin{figure*}
\centerline{\includegraphics[width=19cm]{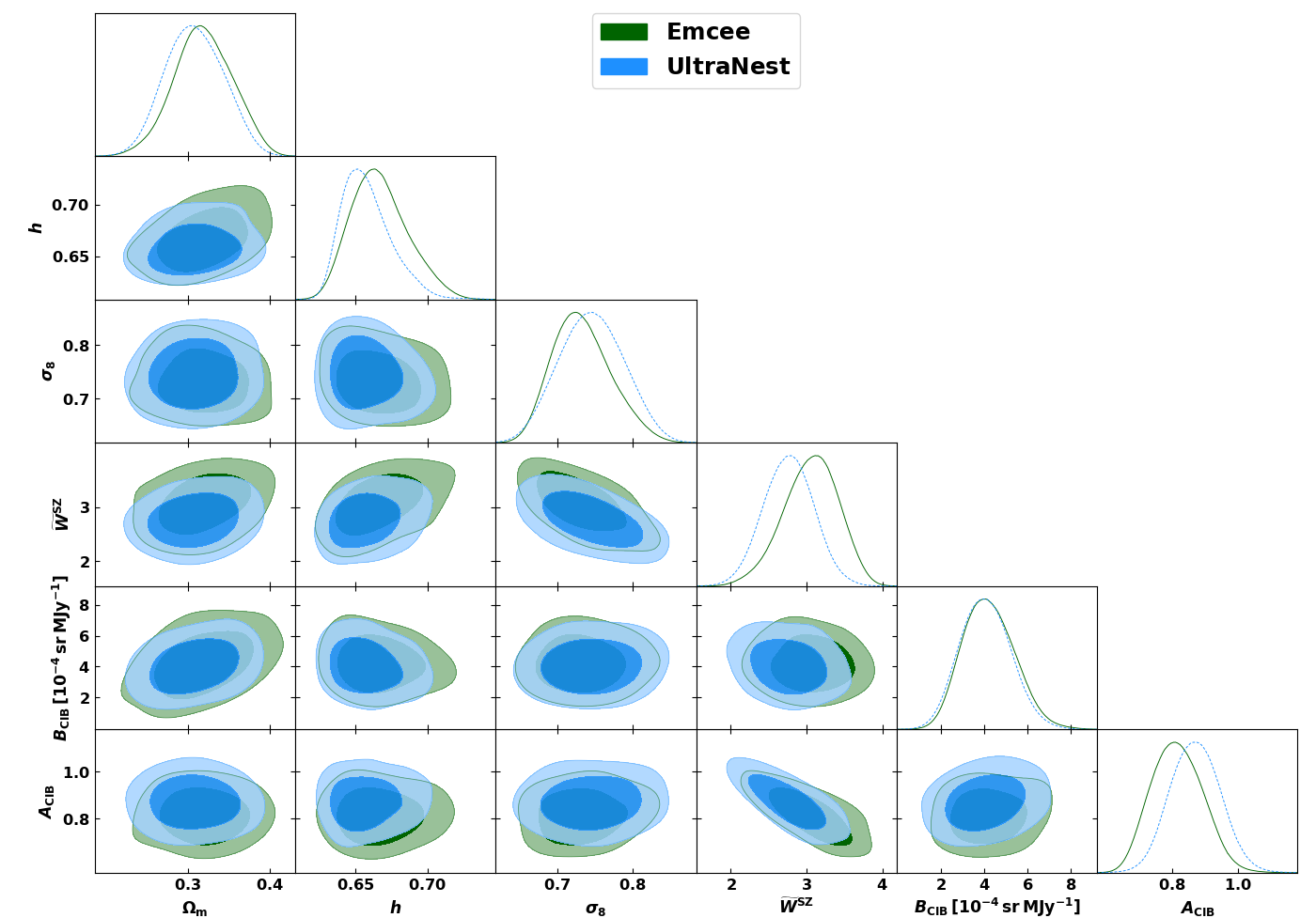}}
\caption{Posterior distributions from MCMC using both \textsc{emcee} and \textsc{ultranest} with the full covariance matrix for the six free parameters in our model. The figures show the joint constraints
for all parameter pairs and the marginalized distributions for each parameter along the table diagonal. Both results are compatible within uncertainties, and they have no obvious discrepancies. Therefore, for this analysis, we present the results obtained using the \textsc{emcee} package, a more versatile and widely employed MCMC engine.}
	\label{fig:compare_constraints}
\end{figure*}

To ensure the accuracy of our parameter estimation, we employ both the \texttt{Python} \textsc{emcee} package~\citep{emcee} package and the \textsc{ultranest} package \citep{ultranest,ultranest2} to explore the parameter space. We used the two packages independently to determine the minimum chi-square or the maximum likelihood for our data and obtained posterior probability distributions on all parameters.
We then used the \textsc{GetDist} \texttt{Python} software detailed in~\citet{Antony19} to plot the final distributions in Figure~\ref{fig:compare_constraints}. One can see from Figure~\ref{fig:compare_constraints} that both packages yield close results in the parameter distributions, therefore we adopt the \texttt{Python} \textsc{emcee} package in our concluding section.

\begin{figure}
	\centering
    \includegraphics[width=3.3in]{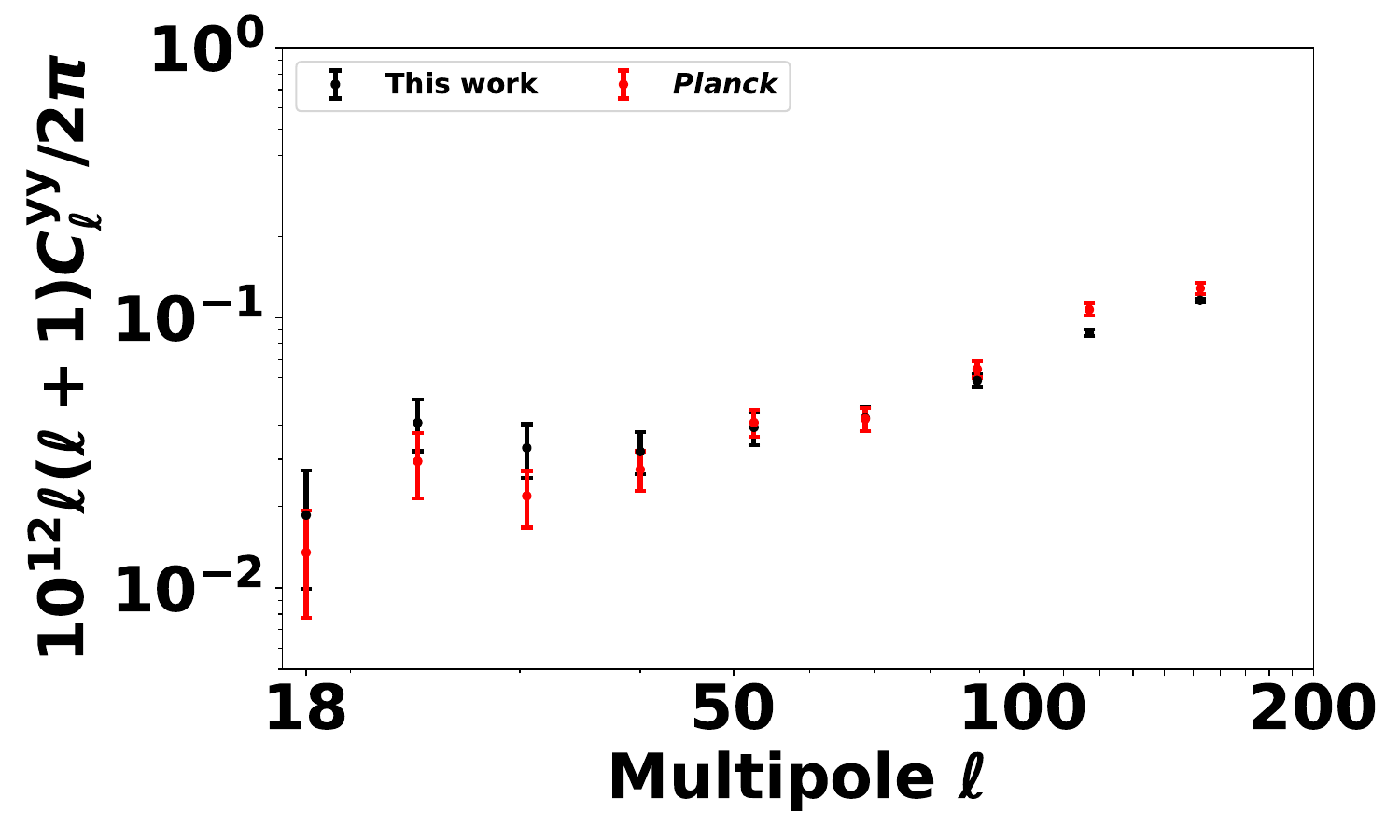}
\caption{Comparison of our $yy$-auto power spectrum with Planck for specific multiples. For both analyses, a 40\% Galactic-plane mask was applied.}
	\label{fig:Compare_yy_with_planck}
\end{figure}

\section{Comparing yy Power Spectrum Analyses}
\label{App:Appendix C}
Here we compare and contrast between the measurements and the modeling presented in our analysis and those presented in \citet{Planck2015_tSZ_paper}. The $yy$ angular power spectrum and the associated error bars measured by Planck were computed using \textsc{XSPECT}~\footnote{XSPECT is a code that was initially written to measure the CMB temperature angular power spectrum from Archeops data. Given a set of maps, it could also measure the cross-correlation angular power spectrum and the associated error bars computed analytically from the data with no MC simulations involved. It utilizes the standard MASTER-like approach \citep{Hivon02} to correct for the beam convolution and the pixelization, as well as the mode-coupling induced by masking sky regions that are foreground-contaminated.}\citep{XSPECT05}. On the other hand, we extracted the $yy$-auto power spectrum from the SZ map in our analysis using \textsc{Namaster}. As shown in Figure~\ref{fig:Compare_yy_with_planck}, one can see that the two measurements on partial-sky are compatible within error bars in the multipole range
$18.0 < \ell < 152.5$. We remind the reader that \citet{Planck2015_tSZ_paper} reported that at multipoles $< 30$, the amplitude of the tSZ power spectrum measured on the MILCA-reconstructed $y$-map is slightly higher than the one measured on the NILC-reconstructed $y$-map. This difference is ascribed to a higher degree of contamination from thermal dust emission at large scales in the MILCA map. 

Now we proceed to compare and contrast the theoretical framework adopted in this work and that adopted in  Appendix A.1 of \citet{Planck2015_tSZ_paper}, to model the observed $yy$ angular power spectrum.

In spherical harmonics, both formalisms agree that the $y$-map is represented by
\begin{eqnarray}
 \label{eq:Spherical_harmonics}
y(\textbf{n}) = \sum_{{\ell} m} y_{{\ell} m}\, Y_{{\ell} m}(\textbf{n}).
\end{eqnarray}  

The tSZ angular power spectrum can then be calculated by carrying out the spherical harmonics decomposition on the sky: 
\begin{eqnarray}
 \label{eq:Spherical_harmonics2}
C^{\rm tSZ}_{\ell} = \frac{1}{2 \ell + 1} \sum_{m} y_{{\ell} m}\, y^{*}_{{\ell} m}
\end{eqnarray}
However, the main difference starts from the assumption upon which the two models are built. While the large-scale bias model assumes a linear biasing relation between the tSZ signal and the underlying matter density field on large scales, the halo model framework assumes that the tSZ angular power spectrum consists of one-halo and two-halo contributions. The one-halo term arises from the Comptonization profile of individual halos within a population that follows a Poisson distribution. The two-halo term, on the other hand, takes into consideration the correlation between individual halos, specifically their two-point correlation function. Hence, while the halo model allows us to model the correlation up to smaller scales, the large-scale bias model provides more details on the large scales instead.

\bibliography{reference}{}
\bibliographystyle{aasjournal}

\label{lastpage}
\end{document}